\documentclass[preprint]{emulateapj}

\usepackage{hyperref}
\usepackage[usenames,dvipsnames]{color}
\hypersetup{colorlinks,citecolor=Blue,linkcolor=Red,urlcolor=Blue}
\usepackage{graphicx}
\usepackage[titletoc]{appendix}
\usepackage{needspace}

\usepackage{amsmath}
 
\begin{document}

\title{The Solar wind prevents re-accretion of debris after Mercury's giant impact} 

\author{Christopher Spalding$^1$ \& Fred C. Adams$^{2,3}$} 
\affil{$^1$Department of Astronomy, Yale University, New Haven, CT 06511}
\affil{$^2$Department of Physics, University of Michigan, Ann Arbor, MI 48109}
\affil{$^3$Department of Astronomy, University of Michigan, Ann Arbor, MI 48109} 
\begin{abstract}
The planet Mercury possesses an anomalously large iron core, and a correspondingly high bulk density. Numerous hypotheses have been proposed in order to explain such a large iron content. A long-standing idea holds that Mercury once possessed a larger silicate mantle which was removed by a giant impact early in the the Solar system's history. A central problem with this idea has been that material ejected from Mercury is typically re-accreted onto the planet after a short ($\sim$\,Myr) timescale. Here, we show that the primordial Solar wind would have provided sufficient drag upon ejected debris to remove them from Mercury-crossing trajectories before re-impacting the planet's surface. Specifically, the young Sun likely possessed a stronger wind, fast rotation and strong magnetic field. Depending upon the time of the giant impact, the ram pressure associated with this wind would push particles outward into the Solar system, or inward toward the Sun, on sub-Myr timescales, depending upon the size of ejected debris. Accordingly, the giant impact hypothesis remains a viable pathway toward the removal of planetary mantles, both on Mercury and extrasolar planets, particularly those close to young stars with strong winds. 
\end{abstract}
\maketitle

\section{Introduction}

The compositions of Mercury, Venus, Earth and Mars, serve as direct windows into the conditions that persisted during the opening 100 million years of our Solar system's formation. Inferred mass ratios between iron and silicates for Venus, Earth and Mars are all roughly consistent with a common, chondritic abundance \citep{Righter2006,Ebel2018}. In contrast, Mercury exhibits an anomalously high iron content. Mercury's iron core possesses a radius exceeding 80\% of the planet's radius, as compared to Earth's more modest 50\%, alongside a correspondingly high bulk density that rivals Earth's, despite the order of magnitude difference in planetary masses \citep{Ash1971}. The explanation for such a disparity in iron content remains mysterious.

Over the decades, many hypothesis have been brought forward to understand Mercury's large core \citep{Solomon2003,Ebel2018}. Broadly, these ideas separate into those that suppose Mercury's high iron-to-silicate ratio is primordial, versus those that propose Mercury formed with a chondritic composition, but that the silicates were subsequently removed, usually by collisions. Primordial scenarios, in contrast, draw upon mechanisms that separate iron and silicates within the proto-planetary material before becoming incorporated into a planet. 

Perhaps the earliest class of ideas supposed that nebular conditions were at such a temperature where silicates were gaseous, but iron had condensed \citep{Urey1951,Lewis1973}. A more recent proposal, still emphasizing Mercury's close-in, high-temperature position in the disk, suggested that the Solar nebular conditions became sufficiently hot at Mercury's orbital distance that Mercury's mantle was vaporized and removed along with the nebular gas \citep{Cameron1985}. In general, temperature-driven mechanisms are inconsistent with modern views of nebular temperature distributions during planet formation \citep{Armitage2011,Hartmann2016}, and appear to contradict the widespread existence of extrasolar planets with periods shorter than Mercury's \citep{Borucki2016}. 


Alternative fractionation pathways exist that do not rely upon temperature, such as photophoresis (an effect related to conductivity; \citealt{Wurm2013}), gas drag (separation of densities; \citealt{Weidenschilling1978}) or magnetic erosion (magnetic attraction of Fe-rich material; \citealt{Hubbard2014}). Whereas these mechanisms may reproduce Mercury's Fe/Si ratio in principle, they typically require special nebular conditions that are either not expected, or contradictory to other findings. Research continues to develop in the field of circumstellar disk chemistry such that primordial fractionation of Si and Fe is not necessarily ruled out. 

In this work, we focus on a separate class of post-formational models that suppose Mercury to have initially conglomerated with a chondritic elemental composition, similar to that of the other terrestrial planets. Subsequently, the silicate mantle was stripped by way of a giant impact \citep{Wetherill1985,Chapman1988,Benz1988,Benz2007}. This model is appealing because an epoch of giant impacts is expected during the latter stages of most envisioned planet formation scenarios \citep{Raymond2004,Raymond2018}. Indeed, giant collisions continue to serve as viable explanations for many other Solar system mysteries, including the formation of the Earth's moon \citep{Hartmann1975,Canup2001,Cuk2012,Lock2016}, the Martian moons \citep{Citron2015,Hyodo2017} the Pluto-Charon system \citep{Canup2005}, and even the tilt of Uranus \citep{Safronov1966,Morbidelli2012}. 

Generally speaking, simulations of Mercury's giant impact have shown that a high-energy impact onto a proto-Mercury of about 2.25 times Mercury's current mass is capable of removing the required mass of silicates to match Mercury's bulk density \citep{Benz1988,Benz2007,Chau2018}. However, a major drawback of the giant impact theory is that much of the material launched from Mercury as vapor tends to condense into solid spherules \citep{Johnson2012} and remain on Mercury-crossing trajectories in the aftermath of the collision. Consequently, the ejected material simply re-accretes over $\sim10\,$Myr timescales \citep{Benz2007,Gladman2009} and the final planetary density is not sufficiently enhanced. 

Proposed solutions to the problem of re-accretion have called upon Poynting-Robertson drag, which acts on a similar timescale to re-accretion for cm-sized particles, the size of debris expected from the condensing ejecta cloud \citep{Burns1979,Benz2007}. However, the high optical depth of the heliocentric ejecta ring can significantly self-shield \citep{Gladman2009} leaving it unclear whether Poynting-Robertson drag alone is enough to prevent re-accretion. This problem has motivated several modifications of the initial impact theory. In one scenario, a larger impactor survives the collision intact, having stripped Mercury's mantle \citep{Asphaug2006,Asphaug2014}. In another, numerous smaller impacts erode Mercury's surface \citep{Svetsov2011}. 

Here, we present the hypothesis that the Solar wind constitutes a robust source of orbital decay for material ejected from Mercury. The Solar wind consists of a nearly radial outflow of ionized particles, travelling outward into the Solar system at 100s of kms$^{-1}$ \citep{Phillips1995}. Any objects in orbit must pass through a non-zero interplanetary density of plasma, inducing a drag force upon their orbits \citep{Burns1979,Mukai1982}. 

Today, the Solar wind-induced drag felt by particles larger than a few microns is roughly 1/3 of that from Poynting-Robertson drag. However if the young Sun possessed a substantially stronger wind than today, the Solar wind drag would increase in turn. It remains difficult to determine the wind properties of stars during the planet-forming period because few direct measurements exist of such early winds \citep{Jardine2018}. Nevertheless winds from Sun-like stars are expected to decay with time, with measurements suggesting 100 times the modern mass-loss rate in younger Solar analogues \citep{Wood2014}. In contrast, the magnitude of Poynting-Robertson drag may only have varied by a factor of a few in the early Solar system, as the Sun's luminosity dropped during the Hayashi track, arriving onto the pre-main sequence slightly dimmer than today \citep{Maeder2008}.

In this work, we show that increasing the Solar wind by a factor of $10-100$ compared to modern leads to removal of ejecta from Mercury within $\sim1$ million year timescales, safely within the 10\,Myr re-accretion time. In sections 2 and 3, we construct a simple, axisymmetric Solar wind model, and compute its impact upon ejected debris in section 4. In section 5, we demonstrate the effect of the Solar wind drag within $N$-body simulations, and in section 6 we discuss the implications of the Solar wind hypothesis.
 \begin{figure*}
\centering
\includegraphics[trim=0cm 0cm 0cm 0cm, clip=true,width=1\textwidth]{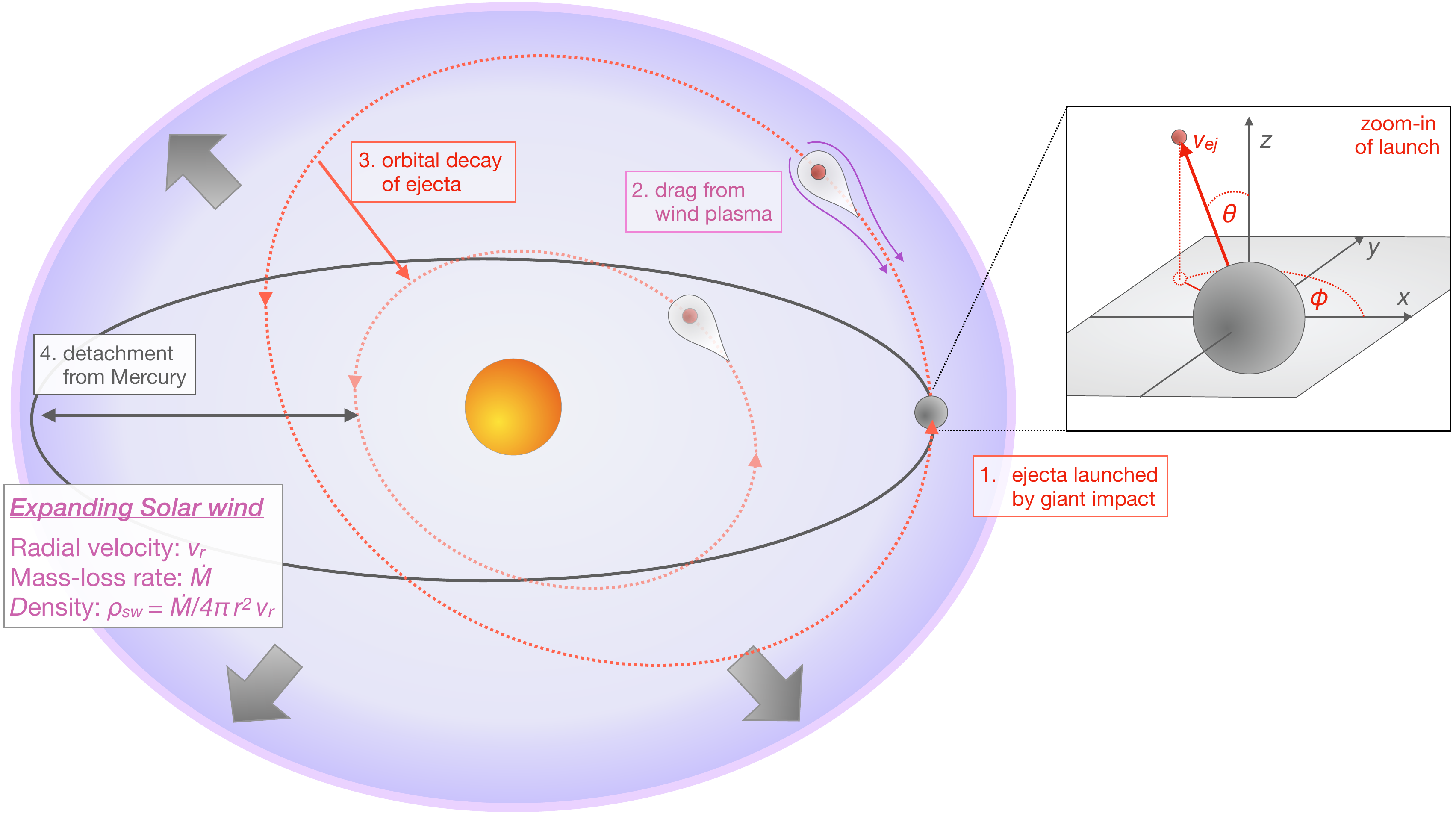}
\caption{Schematic of the physical scenario considered in this work. Material is launched from Mercury (1) subsequent to a giant impact. This material initially follows the outer, red orbit that crosses the orbit of Mercury (grey ellipse). A drag from Solar wind plasma draws angular momentum from the particle's orbit (2), causing the orbit to spiral inward toward the Sun over time (3), and eventually detach from Mercury's orbit (4), inhibiting re-accretion. The zoom-in on the right denotes the coordinate frame used to prescribe the initial trajectories of collisional debris in our $N$-body simulations (Section~\ref{Section5}). The timescale of in-spiral depends grows linearly with particle radius, where a radius of 1\,cm is assumed throughout (see Section~\ref{Size} for a discussion of particle sizes)}
\label{Schematic}
\end{figure*}
\section{Physical setup}

In this section, we outline the physical scenario describing the trajectories of particles ejected from Mercury (see Figure~\ref{Schematic} for a schematic of the problem). We suppose that the precursor of Mercury is impacted by another object, ejecting a large fraction of Mercury's silicate-rich mantle into heliocentric orbit with some velocity $v_{ej}$ \citep{Benz1988,Chapman1988,Gladman2009}. Once in orbit, the material cools and condenses into particles, assumed non-interacting spheres of radius $s$ and density $\rho_s=3$\,gcm$^{-3}$. Each particle follows a Keplerian orbit, of semi-major axis $a$, eccentricity $e$ and velocity \citep{Murray1999}
\begin{align}
\mathbf{v}_K=\sqrt{\frac{GM_\star}{a(1-e^2)}}\big[1+e\cos(f)\big]\bigg[\frac{e\sin(f)}{1+e\cos(f)}\hat{\mathbf{e}}_r+\hat{\mathbf{e}}_\phi\bigg],
\end{align}
where we have defined the true anomaly $f$, stellar mass $M_\star$, gravitational constant $G$ and $\hat{\mathbf{e}}_{r,\phi}$ are unit vectors in the plane of Mercury's orbit (Figure~\ref{Schematic}). 

Subsequent to ejection, the azimuthal ram pressure from the Solar wind plasma leads to orbital evolution of the particles. We construct a model for the young Sun's wind in order to compute whether the Solar wind can remove debris from Mercury-crossing trajectories before re-impacting Mercury's surface \citep{Gladman2009}. For the purposes of the analysis that follows, we assume that this ejecta remains in the plane of Mercury's orbit, thus simplifying the interaction between ejecta and Solar wind plasma. This coplanar assumption will be lifted when we turn to $N$-body simulations in section 5.

The modern Solar wind consists of an ionized plasma expanding into the interplanetary medium with a predominantly radial velocity, which ranges from $\sim400-600$\,kms$^{-1}$ (the slow and fast wind; \citealt{Phillips1995}). The magnitude of mass-loss in the modern Solar system is $\dot{M}_\odot\approx 2\times 10^{-14}M_\odot$year$^{-1}$. 

Mass-loss rates of Sun-like stars appear to decrease with time, with measured values of about 100 times values in Solar analogues at ages of 100s of millions of years \citep{Wood2014}. However, few measurements exist of stellar winds during the first 100 million years subsequent to disk-dispersal. During the disk-hosting stage istelf, star-disk interactions may increase winds to upwards of $\sim10^5$ times modern \citep{White2004,Matt2008,Plavchan2009}, but it is unlikely that such strong wind persist long after the nebular gas has been removed. Accordingly, stellar wind magnitudes during the first 100 million years of planet formation remain mysterious, with estimates ranging from modern levels up to 10,000 times modern \citep{Cohen2014,Fion2018}.  

In addition to the magnitude of mass-loss, young solar analogues generally differ from the Sun by possessing shorter spin periods (ranging between $1-10\,$days; \citealt{Bouvier2014}) and stronger magnetic field strengths ($0.1-1$kGauss; \citealt{Donati2009,Vidotto2014}). The rapidly-rotating magnetosphere would accelerate wind plasma to drive a significant azimuthal component to the wind velocity that we compute below \citep{Lovelace2008,Spalding2018,Carolan2019}. 

A small azimuthal component of the wind's velocity would reduce the net drag upon orbiting debris. However, if the stellar magnetosphere was strong enough, its wind may reach super-Keplerian velocities, causing an acceleration of orbiting material instead of a drag. We suppose that the Solar wind plasma possesses a velocity $\mathbf{v}_{\textrm{sw}}$, this would result in a ram pressure of $C_D\rho_{\textrm{sw}} |\mathbf{v}_{\textrm{sw}}-\mathbf{v}_K|^2$ upon orbiting debris, where $C_D$ is a drag coefficient. We set $C_D=1$ \citep{Mukai1982}, but its exact value depends upon the particle's composition and size.

The above ram pressure leads to a force of
\begin{align}\label{force}
\mathbf{F}&=C_D\pi s^2\rho_{\textrm{sw}}|\mathbf{v}_{\textrm{sw}}-\mathbf{v}_p|\big(\mathbf{v}_{\textrm{sw}}-\mathbf{v}_K\big),
\end{align} 
where the wind-facing surface area of the particles is taken as $\pi s^2$. We may obtain a first order approximation to $\mathbf{F}$ by assuming null eccentricity, together with considering the limits where the azimuthal component of the wind is negligible (as it is today) and the wind's radial velocity $v_r$ greatly exceeds the Keplerian velocity. Finally, we approximate the wind as spherical to write the plasma density as
 \begin{align}
 \rho_{\textrm{sw}}\approx \frac{\dot{M}}{4\pi r^2 v_r}.
 \end{align}
 With the above simplifications, the drag given by Equation~(\ref{force}) may be used to compute a timescale of orbital decay \citep{Spalding2018}:
\begin{align}\label{tdrag}
\tau_a&\equiv \frac{a}{\dot{a}}\approx\frac{8\pi\rho_s\,sa^2}{3C_D\dot{M}},
\end{align}
where $\dot{M}$ is the mass-loss rate of the star.

Notice that under the current level of approximation, the drift timescale is independent of the wind's velocity. The modern Sun's mass-loss is $\dot{M}=\dot{M}_\odot\approx 2\times10^{-14}$M$_\odot$yr$^{-1}$ \citep{Phillips1995}, but as mentioned above was likely exceeded by orders of magnitude during terrestrial planet formation. Drift timescales are proportional to particle radius $s$, which is likely to vary with impactor parameters and the composition of debris \citep{Melosh1991,Benz2007,Johnson2012,Chau2018}. By combining Smoothed-Particle Hydrodynamics (SPH) simulations with thermodynamic computations, \citet{Benz2007} calculated that ejected particles possess typical sizes of $s\sim1$\,cm and it is this value that we assume throughout. However, we return to this problem in Section~\ref{Size} to discuss avenues toward updated estimates.

\begin{figure}[h!]
\centering
\includegraphics[trim=0cm 0cm 0cm 0cm, clip=true,width=1\columnwidth]{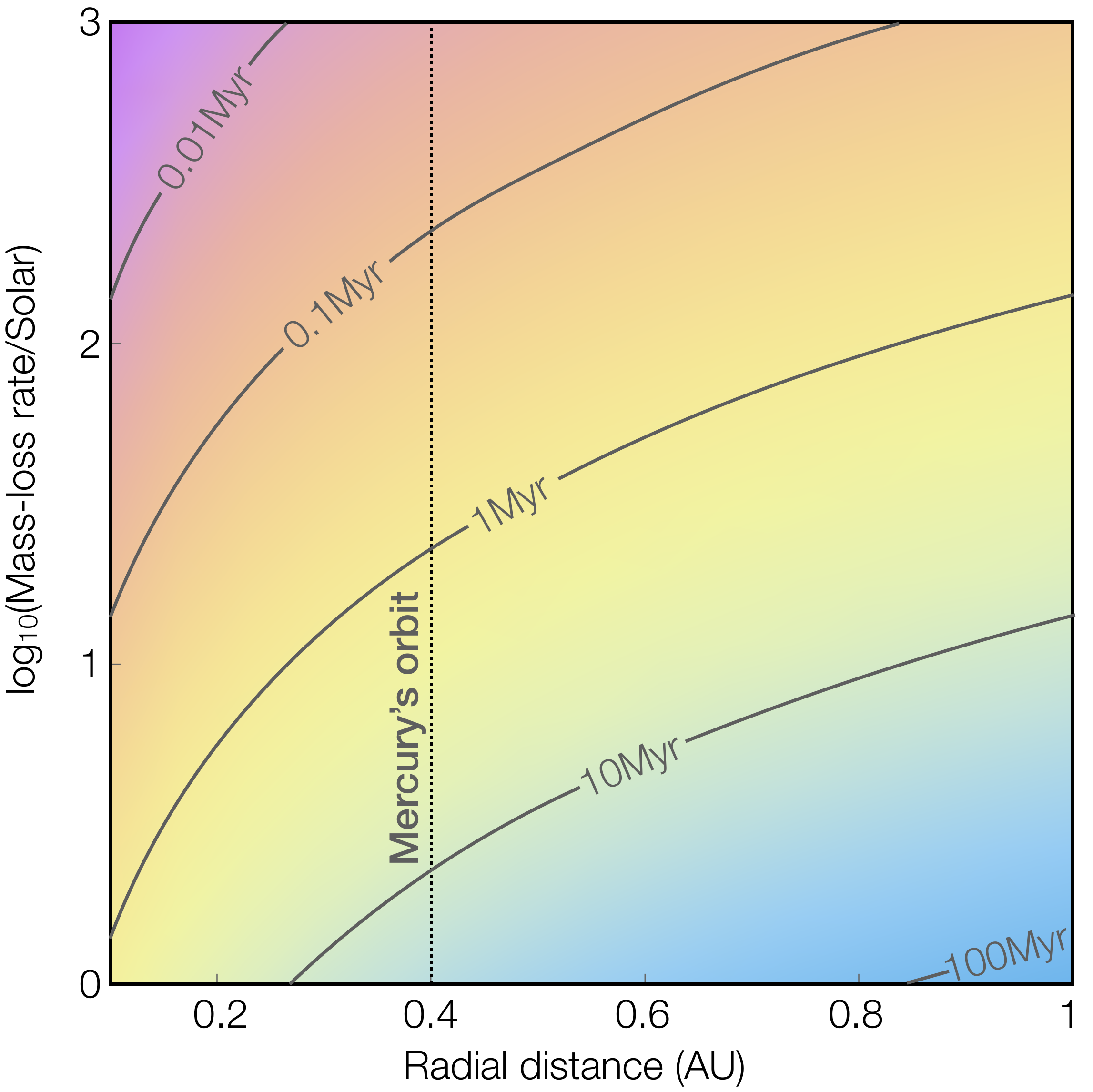}
\caption{Contours of orbital migration timescales in the case of a purely radial wind ($v_\phi=0$), where background colors depict the gradation in drift timescales. The Solar wind magnitude is given on the $y$-axis in units of the modern-day Sun's mass-loss. The orbit of Mercury ($\sim0.4$\,AU) is denoted by a dashed line, and a 1\,Myr drift timescale contour is marked in grey. Values of mass-loss of $\gtrsim10$ times modern are sufficient to cause orbital migration of 1cm particles over a 1\,Myr timescale, preventing re-accretion of ejecta onto Mercury's surface.}
\label{Vr}
\end{figure}

Given the parameters above, we find that the drift timescale takes the magnitude
\begin{align}
\tau_a&\sim 0.2 \bigg(\frac{a}{0.4\textrm{AU}}\bigg)^2\bigg(\frac{s}{1\textrm{cm}}\bigg)\bigg(\frac{\dot{M}}{100\dot{M}_\odot}\bigg)^{-1}\textrm{Myr},
\end{align}
which is plotted for various orbital radii in Figure~\ref{Vr}. As can be seen, Solar winds exceeding roughly 10--20 times the present magnitude are sufficient to cause outward migration of cm-sized particles over Myr timescales. The re-accretion timescale of ejecta found previously is on the order of 1--10\,Myr \citep{Gladman2009}. Thus, from the simple treatment above, the Solar wind constituted a significant component to the orbital evolution of material ejected from the primordial Mercury subsequent to a purported giant collision.

Despite the promising results of timescale considerations, ejecta from Mercury are unlikely to be launched onto circular, heliocentric orbits. Moreover, we have not yet included the effect of the azimuthal component of the wind velocity, which may increase or decrease drift timescales depending upon the magnitude and sign of the quantity $\mathbf{v}_{\textrm{sw}}-\mathbf{v}_K$.

\section{Stellar wind model}

One of the earliest quantitative models of the Solar wind considered a 1-dimensional, steady-state approximation of an isothermal gas moving radially-outward from an un-magnetized, non-rotating star \citep{Parker1965}. Such a model is remarkably efficient at recreating the radial velocity of the Solar wind, but is incapable of deriving an azimuthal velocity. A 2-dimensional, axisymmetric model that takes rotation and magnetism into account was developed soon after \citep{Weber1967}. Such a ``Weber-Davis" model assumed that the wind consisted of an infinitely-conductive plasma that interacted with the star's rotating magnetosphere through the principles of ideal magnetohydrodynamics. 

In the intervening decades, Solar wind models have continuously grown in sophistication, including fully 3-dimensional magnetohydrodynamics (MHD) simulations \citep{Cranmer2011,Cohen2014,vanderHolst2014}. Our primary goal for this work is to compute an approximate radial profile of the wind's azimuthal velocity, given a stellar mass-loss rate, magnetic field and spin rate. Thus, the Weber-Davis model will suffice for our purposes, but we draw the evolution of $\dot{M}$, magnetic field and spin rate from published observations \citep{Vidotto2014} and 3-dimensional MHD simulations \citep{Fion2018}.

\subsection{Weber-Davis model}

The picture of the Solar wind we employ considers the wind plasma to expand radially outwards along the stellar equator. Its outward motion is governed by both pressure and magnetic stresses, but its azimuthal acceleration is dominated by magnetic torques. A pathway toward a solution of this ``Weber-Davis" model has been presented in numerous works \citep{Weber1967,Hartmann1982,Lamers1999} and so we only provide a brief outline of the solution here. Closely following \citet{Hartmann1982}, we assume that both the magnetic field $\mathbf{B}$ and wind velocity are axisymmetric, and possess only radial ($v_r,\,B_r$) and azimuthal ($v_\phi,\,B_\phi$) components. The condition $\nabla \cdot \mathbf{B}=0$ in spherical coordinates demands that
\begin{align}
B_r=B_{r,0}\bigg(\frac{R_\star}{r}\bigg)^{2},
\end{align}
where $B_{r,0}$ is the magnetic field strength at the stellar photosphere, taken to be when $r=R_\star$, the stellar radius. Note that all calculations are carried out in the equatorial plane.

If Faraday's law is combined with the assumptions of ideal MHD (i.e., infinite conductivity), we obtain a steady-state relationship of
\begin{align}
B_\phi=B_r\Bigg[\frac{v_\phi-\Omega_\star r}{v_r}\Bigg],
\end{align}
where the stellar angular velocity $\Omega_\star$ has been introduced. For simplicity, we assume that the wind is isothermal, with isothermal sound speed $c_s=(k T/m_H \mu)^{1/2}$ given in terms of the photosphere temperature $T=10^6$\,K, the mean molecular number $\mu=0.6$, the mass of a proton $m_H$ and the Boltzmann constant $k$ \citep{Lamers1999}. 

Each parcel of plasma carries with it a constant specific angular momentum $L$ and energy $E$, which are derived from the steady-state, azimuthal and radial components of the momentum equation respectively:
\begin{align}
r:\,\,\,\,\,\,\,\,0&=\frac{1}{2}\frac{d}{dr}\bigg[v_r^2+v_\phi^2\bigg]+c_s^2\frac{d}{dr}\big[\ln(\rho_{sw})\big]+\frac{GM_\star}{r^2}\nonumber\\
&-\frac{v_\phi}{r}\frac{d}{dr}\big[r v_\phi\big]+\frac{B_\phi}{\mu_0 \rho r}\frac{d}{dr}\big[rB_r\big]\\
\phi:\,\,\,\,\,\,\,\,0&=\frac{B_r}{\mu_0}\frac{d}{dr}\big[rB_\phi\big]-\rho v_r\frac{d}{dr}\big[rv_\phi\big],
\end{align}
where $\mu_0=4\pi\times10^{-7}$\,Hm$^{-1}$ is the vacuum permeability. 

The conserved angular momentum is extracted via integration as
\begin{align}
L=rv_\phi-\Bigg[\frac{rB_rB_\phi}{\mu_0 \rho v_r}\Bigg]=r_A^2 \Omega_\star,
\end{align}
where the Alfv\'en radius $r_A$ is defined as the point where the radial wind velocity is equal to the Alfv\'en velocity, defined as $v_{A,r}\equiv B_r/\sqrt{\mu_0 \rho_{sw}}$ \citep{Lamers1999,Bellan2008}. Likewise, integration yields the conserved energy
\begin{align}
E\equiv \frac{1}{2}\big[v_r^2+v_\phi^2\big]+c_s^2 \ln(\rho_{sw})-\frac{GM_\star}{r}-\frac{r\Omega_\star B_r B_\phi}{\mu_0 \rho_{sw} v_r}.
\end{align}
Finally, rearranging the momentum equation yields to the relationship
\begin{align}\label{continue}
\frac{r}{v_r}\frac{dv_r}{dr}=\frac{\big(v_r^2-v_{A,r}^2\big)\big(2c_s^2+v_\phi^2-{GM_\star/r}\big)+2v_rv_\phi v_{A,r}v_{A,\phi}}{\big(v_r^2-v_{A,r}^2\big)\big(v_r^2-c_s^2\big)-v_r^2v_{A,\phi}^2},
\end{align}
where we define $v_{A,\phi}\equiv B_\phi/\sqrt{\mu_0 \rho_{sw}}$.

By inspection of Equation~(\ref{continue}), any solution passing through a point where the denominator vanishes must also cause the numerator to vanish. This situation occurs at 2 radii, which are $r_f$ (the fast point) and $r_s$ (the slow point), and physically correspond to the transition of the radial wind velocity through the fast and slow magnetosonic points \citep{Weber1967,Hartmann1982,Lamers1999,Bellan2008}. 

The requirement of vanishing numerator and denominator at both points constitutes a system of 4 simultaneous equations. Two further equations arise from the constancy of the energy $E$ at the stellar surface $r=R_\star$ and at the Alfv\'en radius $r=r_A$. Altogether, these 6 equations must be solved for 6 unknowns, which include the 3 radii $\{r_s,r_f,r_A\}$, the values of radial wind velocity at $r_f$ and $r_s$, and finally the velocity of the wind at the stellar photosphere. We obtain these six parameters by way of a Newton-Rhapson solver, from which the Solar wind properties, and thus the ejecta drift timescales, may be obtained as a function of radius. 

 \begin{figure}[h!]
\centering
\includegraphics[trim=0cm 0cm 0cm 0cm, clip=true,width=1\columnwidth]{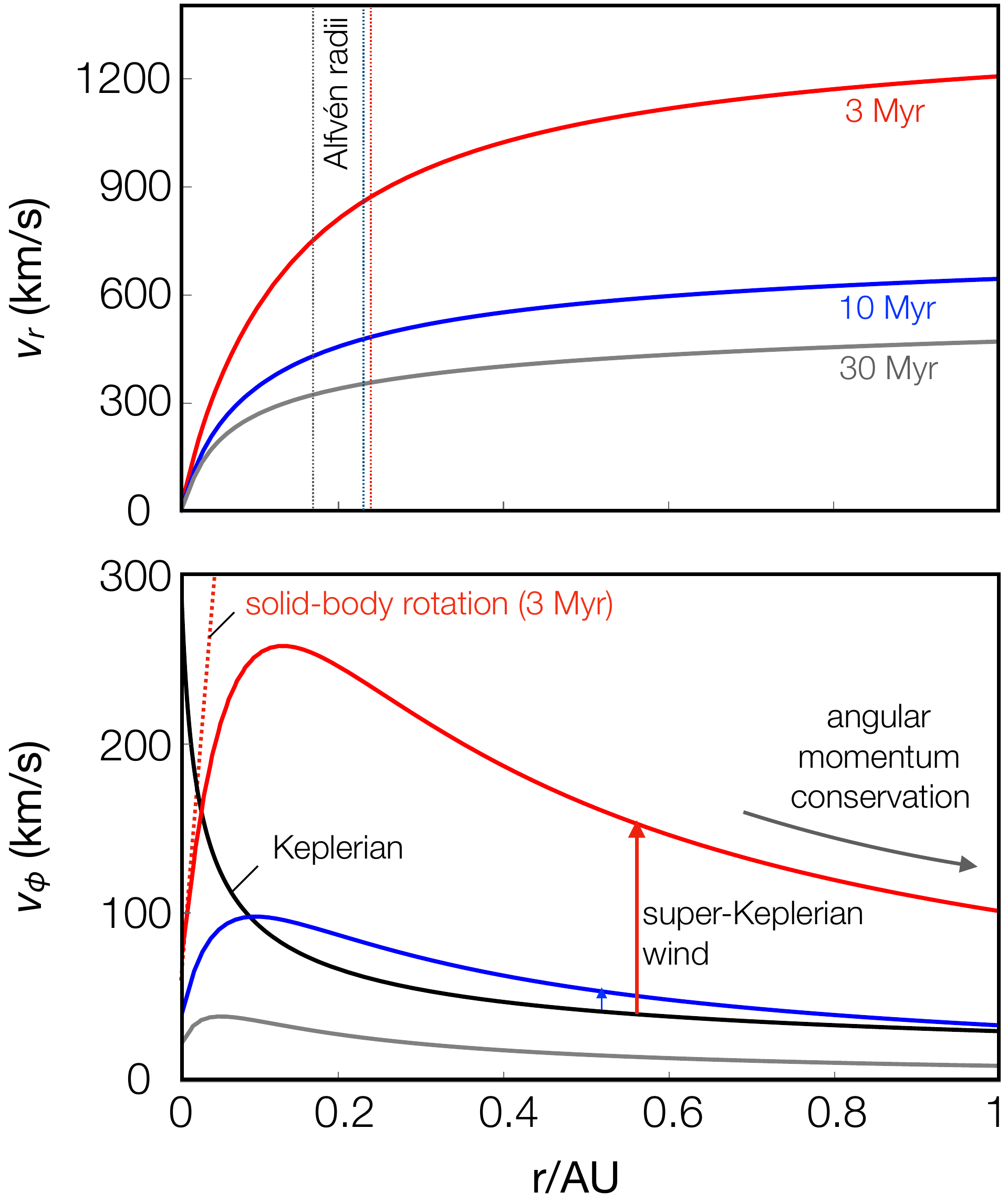}
\caption{The radial (top) and azimuthal (bottom) velocities of the Solar wind as computed at 3 different times--3\,Myr, 10\,My and 30\,Myr. The azimuthal velocity is compared to the Keplerian orbital velocity in order to illustrate the regions and times where the Solar wind acts to accelerate particles (where $v_\phi$ exceeds Keplerian). Subsequent to 30\,Myr, the azimuthal velocity drops sufficiently below both the Keplerian and radial velocities, eventually becoming negligible. At all times, the azimuthal velocity rises steadily from the stellar surface, eventually falling off at large distance owing to the weakening magnetic stresses \citep{Weber1967}.}
\label{vPhi}
\end{figure}

Illustrative radial profiles of $v_r$ and $v_\phi$ are presented in Figure~\ref{vPhi} at three epochs--3, 10 and 30\,Myr. The radial velocity increases monotonically with distance, reaching a constant as $r\rightarrow \infty$. In contrast, the azimuthal velocity displays an extremum in velocity slightly interior to the Alfv\'en radius (vertical dotted lines). Physically, this peak arises because close to the star the magnetic field is strong enough to accelerate an outwardly-expanding ring of plasma over a timescale that is short compared to the radial expansion timescale of the plasma. Approximately, this strong coupling locks plasma to solid-body rotation close to the star's surface (denoted by the slanting, dashed red line). At larger distances ($r>r_A$), magnetic coupling weakens, such that the expanding ring of plasma no longer receives significant acceleration from the weakening magnetic stresses. In order to approximately conserve its angular momentum $v_{\phi}$ approximately falls as $1/r$ \citep{Carolan2019}.

At times earlier than about 10\,Myr, a substantial range of radii exist where $v_\phi$ exceeds the Keplerian velocity (the black line in the bottom panel). We discuss the significance of this in greater detail below, but qualitatively it suggests that debris occupying Mercury's orbit at earlier times are more likely to experience orbital expansion than at later epochs, when orbital contraction is favored. 

Up until recently, in situ measurements of the Solar wind's azimuthal velocity were unavailable at distances comparable to Mercury's orbit, and so the theoretically-expected peak in $v_\phi$ could not be tested. However, the Parker Solar Probe recently detected azimuthal winds, increasing toward the Sun at distances inward of 0.2\,AU \citep{Kasper2019}. Surprisingly, these winds possessed azimuthal components of up to 40\,kms$^{-1}$, at least an order of magnitude above those predicted for the Sun at its current age. Indeed, these speeds are more consistent with the winds we compute at an age of 30\,Myr (bottom panel of Figure~\ref{vPhi}). It is impossible to tell as of yet how these theoretical underestimations propagate to younger, more active stars, and thus whether or not our calculations serve as lower limits of the early wind strength. Undoubtedly, the upcoming closer passes of the Solar Probe to the Sun will help shed light on this unfolding mystery.
\needspace{5\baselineskip}
\subsection{Early stellar properties}

The epoch of giant impacts persisted throughout approximately the first 100 million years of Solar system history \citep{Raymond2018}, though the exact time of Mercury's hypothesized impact is uncertain (see Section~\ref{timing}). As mentioned above, young stars typically possess spin periods between 1 and 10 days and magnetic fields between 0.1 and 1 kGauss \citep{Bouvier2014,Vidotto2014}. A less well-constrained aspect lies in choosing the mass-loss rate $\dot{M}$, alongside its time-dependence. Given the lack of direct measurements, previous work has typically resorted to 3-D magnetohydrodynamic models from which the mass-loss emerges as an output of the model \citep{Cohen2014,Fion2018}. 

Unfortunately, these computational approaches do not approach a consensus, with estimates varying widely, though generally the period $P_\star$, $\dot{M}$ and $B_\star$ all decrease with time. We adopt the time-scalings presented in \citet{Fion2018},
\begin{align}
\dot{M}(t)&=\dot{M}_0\Bigg(\frac{t}{\textrm{Myr}}\Bigg)^{-3/4}\nonumber\\
B_\star(t)&=B_0\Bigg(\frac{t}{\textrm{Myr}}\Bigg)^{-1/2}\nonumber\\
P_\star(t)&=P_0\Bigg(\frac{t}{\textrm{Myr}}\Bigg)^{1/2},
\end{align}
with magnitudes that are chosen to match observed T Tauri magnetic fields and spin periods, as well as allowing the wind to pass through 100$\dot{M}_\odot$ during the period of terrestrial planet formation:
\begin{align}
\dot{M}_0=10^3\dot{M}_\odot\,\,\,\,\,\,B_0=0.1\,\textrm{kG}\,\,\,\,\,\,P_0=1\,\textrm{day}.
\end{align}
Though the uncertainty in the above parameters is large, it is important to point out that the time at which Mercury's purported giant impact occurred is currently unconstrained. Thus, even if an exact knowledge existed of the young Sun's stellar parameters, the conditions persisting during Mercury's giant impact would remain with large uncertainties. Accordingly, our goal is to explore whether a broad range of stellar parameters and impact times are consistent with removal of ejected material.

In addition to the above parameters, the stellar radius itself would be changing throughout the terrestrial planet formation period. Specifically, the Sun contracts along the fully-convective Hayashi track, entering the Henyey track after several tens of Myr \citep{Shu1987,Maeder2008}. Using stellar models from the online database of \citet{Siess2000}, we computed the time evolution of a Sun-like star's radius. We find that this contraction is well-approximated by the equation
\begin{align}
R_\star(t)=R_\odot\Bigg[0.9+\exp\bigg(-\frac{t}{10\,\textrm{Myr}}\bigg)\Bigg],
\end{align}
where $R_\odot$ is the present-day Sun's radius. The decay of the stellar radius provides a significant reduction in the effect of the wind, because of the $R_\star^2/r^2$ dependence of $B_r$. 

With the above prescriptions, the stellar wind properties as a function of $r$ and $t$ are fully specified. We now turn to the influence this wind has upon material ejected from Mercury's surface. 

\section{Orbital evolution of ejecta}

In the previous section, we obtained a prescription for the Solar wind as a function of heliocentric distance and time. Here we use these results to compute the orbital evolution of ejecta launched from Mercury. The drag $\mathbf{F}$ experienced by ejecta is given in vector form by Equation~(\ref{force}). There exist analytic expressions that relate the radial $F_r$ and azimuthal $F_\phi$ components of this drag to the time-evolution of a particle's eccentricity $e$ and semi-major axis $a$, which take the form \citep{Burns1976,Hedman2018}:
\begin{align}
\frac{da}{dt}&=2\sqrt{\frac{a^3}{G M(1-e^2)}}\Bigg[\frac{F_r}{m_p} e\sin(f)+\frac{F_\phi}{m_p}\big[1+e\cos(f)\big]\Bigg]\nonumber\\
\frac{de}{dt}&=2\sqrt{\frac{a}{G M(1-e^2)}}\Bigg[\frac{F_r}{m_p} \sin(f)+\frac{F_\phi}{m_p}\bigg(\frac{e+\cos(f)}{1+e\cos(f)}\bigg)\Bigg].
\end{align}
We assume that these forces act adiabatically, i.e., that $\|\dot{e}/e|\ll n$ where $n$ is the orbital mean motion. Then the appropriate evolutionary equations emerge from an average over time. A time-average may be achieved by using Kepler's second law $r^2 \dot{f}=$\,constant, such that \citep{Burns1979}
\begin{align}
\langle\dot{X}\rangle\equiv\frac{1}{P}\int_0^P\dot{X}dt=\frac{1}{2\pi a^2\sqrt{1-e^2}}\int_0^{2\pi}\dot{X}r^2 df,
\end{align}
for the orbital parameter $X$, and $r=a(1-e^2)/(1+e\cos(f))$. 
 \begin{figure*}
\centering
\includegraphics[trim=0cm 0cm 0cm 0cm, clip=true,width=1\textwidth]{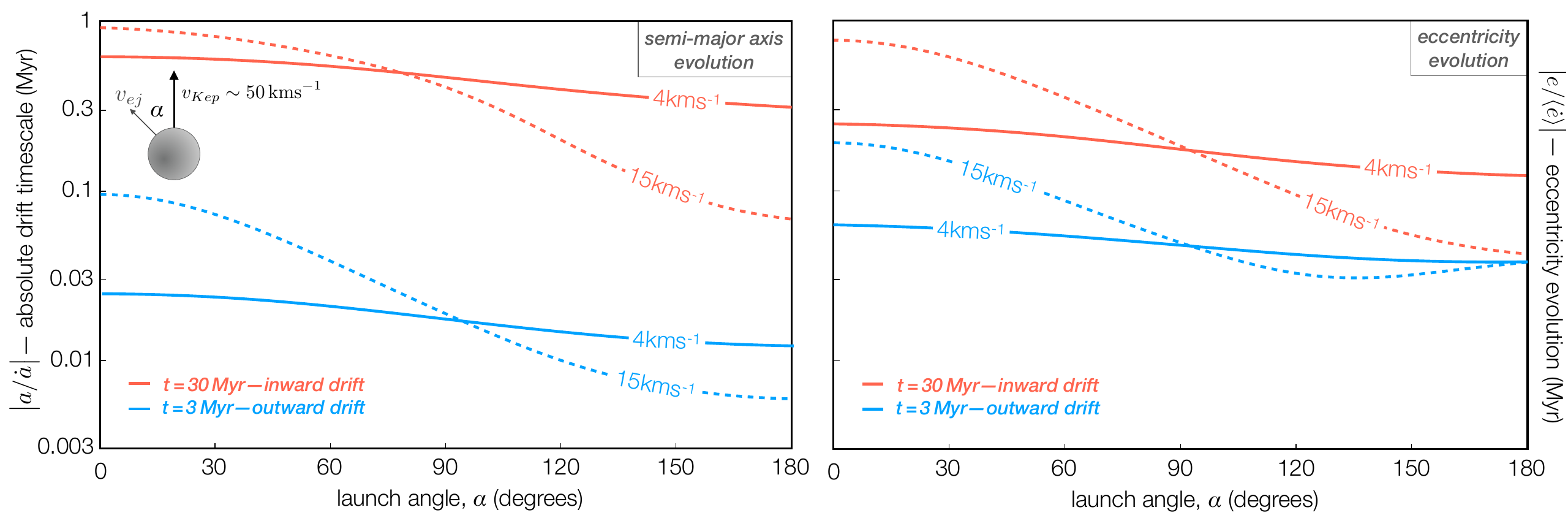}
\caption{The Solar wind-induced evolutionary timescales of semi-major axis (left) and eccentricity (right) as a function of the launch angle relative to Mercury's orbit. We depict 2 launch velocities (4km/s and 15km/s) at an early (3\,Myr) and late (30\,Myr) stage. For all parameters, evolutionary timescales are shorter than 1\,Myr, indicating efficient removal before re-accretion.}
\label{angles}
\end{figure*}

 Unfortunately, the above is not a trivial computation when particles possess significant eccentricities, which is likely to be the case appropriate for ejecta flung from Mercury into heliocentric orbit \citep{Gladman2009}. To illustrate this, suppose that a particle is ejected with velocity $v_{ej}$ relative to Mercury, directed in Mercury's orbital plane but at an angle $\alpha$ to Mercury's orbital velocity. For simplicity, we assume Mercury's orbit to be circular, such that the eccentricity and semi-major axis of the particle are given by
 \begin{align}
 a&=\frac{a_M}{1-2\beta \cos(\alpha)-\beta^2}\nonumber\\
 e^2&=1-\big(1-\beta^2-2\beta \cos(\alpha)\big)\big(1+\beta\cos(\alpha)\big)^2,
 \end{align}
 where $\beta\equiv v_{ej}/v_{Kep}$ is the normalized relative ejecta velocity. 
 
 Using the relationships above, we may write down a formula for the orbital and eccentricity decay time as a function of the particle's ejected velocity and direction:
 \begin{align}\label{driftTime}
 \tau_e\equiv \frac{e}{\langle \dot{e} \rangle}\bigg|_{\beta,\alpha}\,\,\,\,\,\,\,\,\,\,\,\,\tau_a\equiv \frac{a}{\langle \dot{a} \rangle}\bigg|_{\beta,\alpha}.
 \end{align}
 These timescales are illustrated in Figure~\ref{angles} in order to depict the dependence of drift and eccentricity decay timescale upon the velocity of ejection. 
 
As illustrated in Figure~\ref{angles}, the eccentricity damping timescale is similar in both sign and magnitude to $\tau_a$. Interestingly, this means that during the first few Myrs, when azimuthal wind velocities are high, semi-major axes and eccentricities of debris both grow at similar rates. At later times, they both damp at similar rates. We do not explore the eccentricity evolution in great detail here, but it serves as a potentially important dynamical influence upon debris disks around young stars.
 
\begin{figure}[h!]
\centering
\includegraphics[trim=0cm 0cm 0cm 0cm, clip=true,width=1\columnwidth]{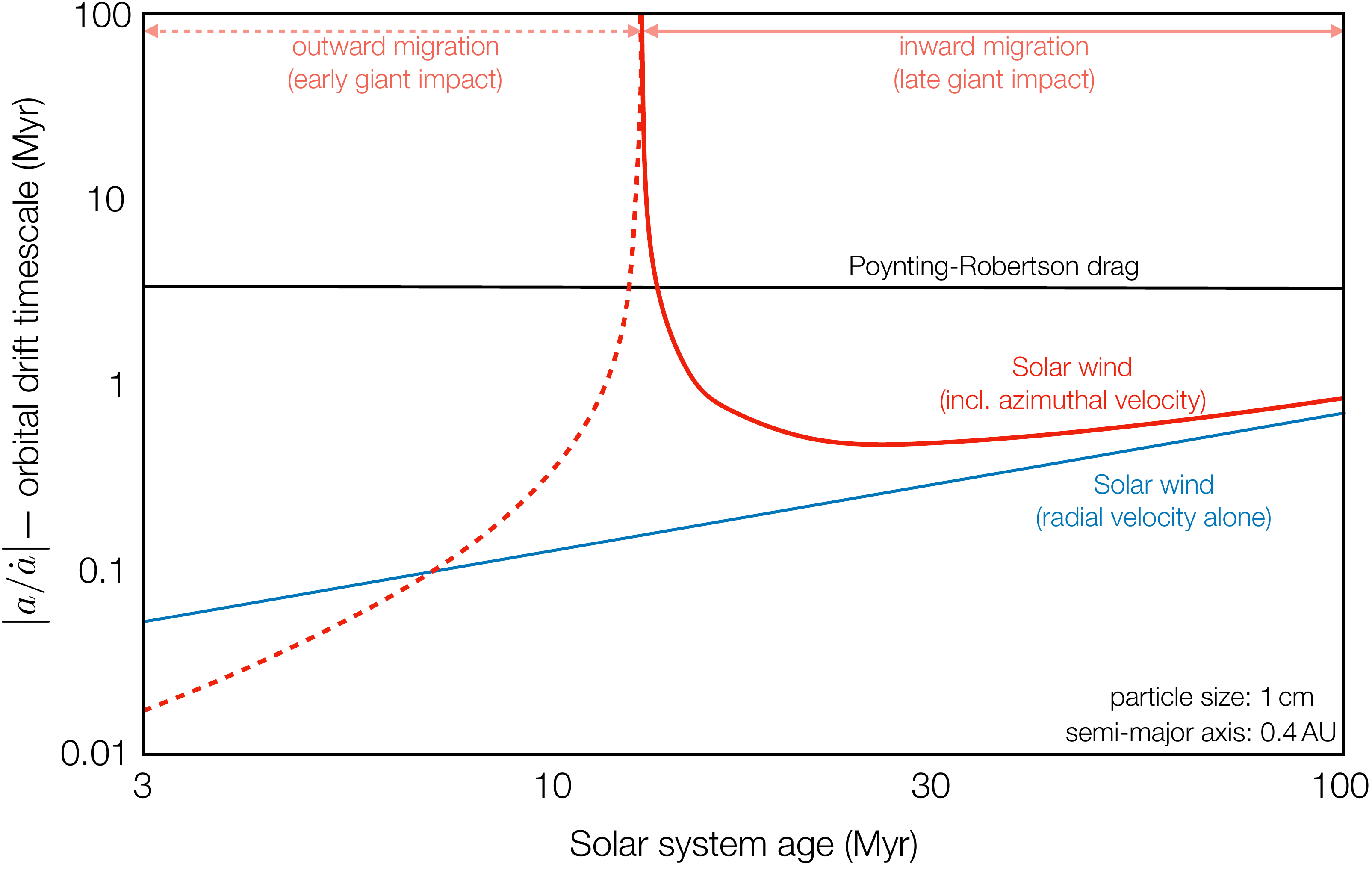}
\caption{Orbital evolution timescales of a 1 cm sized particle under the action of the Solar wind resulting from the solution of the Weber-Davis model (red)--where dotted lines denote times of outward migration and solid lines denote times of inward migration (orbital decay). Also plotted are the drag timescales associated with a purely radial Solar Wind (blue) and from Poynting-Robertson drag (black).}
\label{DragCompare}
\end{figure}

 In Figure~\ref{DragCompare}, we set $e=0$ and plot the orbital evolution timescale owing to wind-induced drag at Mercury's orbit as a function of the age of the system (beginning at 3\,Myr). For comparison, we include the timescale associated with Poynting-Robertson drag, which takes the form \citep{Burns1979,Gladman2009}
\begin{align}
\tau_{PR}&\approx\frac{8\pi a^2 c^2 \rho_s s}{3L_\star}\nonumber\\
&\approx 7\,\textrm{Myr} \bigg(\frac{s}{\textrm{cm}}\bigg)\bigg(\frac{a}{0.4\,\textrm{AU}}\bigg)^2,
\end{align}
in terms of the Sun's luminosity $L=3.8\times 10^{26}$\,W and the speed of light $c=3\times10^8$\,ms$^{-1}$. As mentioned above, the Sun's luminosity will slightly exceed the modern value at 3\,Myr and drop slightly below it at 100\,Myr \citep{Siess2000,Maeder2008}. Given the relatively small magnitude of such a variation we simply use the Sun's current luminosity for comparison. 

The drift timescale associated with the Solar wind is significantly shorter than that associated with Poynting-Robertson drag over most of the wind's evolution. However, it is worth noting that between about 10 and 20\,Myr, the wind-induced timescale grows substantially. This occurs because the wind transitions from a super-Keplerian to a sub-Keplerian configuration, which passes briefly through a phase where the relative velocity between debris and wind is small.
 
 Most importantly for our work, the magnitude of $\tau_a$ is consistently shorter than 1\,Myr across a wide range of launch parameters, dropping below 0.2\,Myr for low launch velocities. Accordingly, over the entire feasible range of launch velocities, ejecta is removed over a timescale shorter than the 10\,Myr reaccretion time presented previously \citep{Gladman2009}.
 
\section{Numerical Simulations}\label{Section5}

In the quantitative treatment above, we concluded that the orbital evolution timescales for cm-sized particles ejected from the surface of Mercury typically lie in range of $0.1-1$\,Myr. This semi-analytic treatment suggests that the ejecta would be removed from Mercury-crossing trajectories before re-accreting. However, perturbations from Mercury itself, as well as from the other planets complicate this picture beyond the above scenario of a single particle orbiting the Sun. Secular and mean motion resonances permeate the Solar system, potentially affecting the evolution of debris \citep{Gladman1996}. 

In order to test our analytic results within a more realistic early Solar system scenario, we perform $N$-body simulations of the ejecta orbits, subject to the gravitational forces of Mercury, Venus, Earth, Mars, Jupiter and the Sun. We carry out simulations, both with and without the Solar wind-induced drag in order to compute the efficiency with which the Solar wind may remove debris from Mercury's vicinity. 
\needspace{5\baselineskip}
\needspace{5\baselineskip}
\subsection{Computational set up}

All simulations were carried out using the \textit{REBOUND} symplectic integration package, where the hybrid integration scheme `Mercurius' was employed \citep{Rein2012,Rein2015}. This integration scheme employs a WHFast algorithm \citep{Wisdom1991}, where our chosen timestep was $0.2\,$days, transitioning to an IAS15 algorithm to deal with close encounters \citep{Rein2014}. Multiple timesteps were checked, ranging from 2 days to 0.2 days, but there was no systematic dependence observed between collision rate and step size. In order to model Solar wind-induced drag, we prescribed exponential evolution of both the eccentricities and semi-major axes of ejecta particles by way of the additional forces options within the \textit{REBOUNDx} package \citep{Tamayo2019}. 

We began with a control simulation that did not include the Solar wind, similarly to \citet{Gladman2009}. These integrations were performed for 10 million years each, and 2 particle ejection velocities (the velocity of the ejecta relative to Mercury) were chosen, $v_{ej}=\{4,\,10\}\,$kms$^{-1}$. In parallel, we simulated these two launch velocities under the action of the Solar wind. In these simulations, eccentricities and semi-major axes ejecta orbits were forced to exponentially decay over respective timescales of $\tau_e$ and $\tau_a$, computed according to Equations~(\ref{driftTime}). 

Given that eccentricity-damping timescales computed above are relatively fast, and for the sake of computational speed, we employed Equations~(\ref{driftTime}) to lowest order in eccentricity, such that
\begin{align}
2\tau_e=\tau_a=\frac{8\pi\rho_s\,sa^2}{3C_D\dot{M}},
\end{align}
and assumed that $s=1\,$cm, the typical size of debris computed by \citet{Benz2007}. Accordingly, the drift times are
\begin{align}
2\tau_e=\tau_a\approx 0.23 \bigg(\frac{a}{0.4\textrm{AU}}\bigg)^2\bigg(\frac{100\dot{M}_\odot}{\dot{M}}\bigg)\,\textrm{Myr},
\end{align}
and were updated with the particles' new semi-major axes after each $1000$\,year interval throughout the simulation. We simulated two cases for each of $v_{ej}=4$\,kms$^{-1}$ and $v_{ej}=10$\,kms$^{-1}$, with $\dot{M}=\{10,\,100\}M_\odot$. Cumulatively, we ran 3 cases for each of the 2 launch velocities, with these 3 cases denoted as ``weak wind", ``strong wind", and ``wind-free" in Figure~\ref{NumSim}.
\subsection{Initial conditions}

Each simulation began with 110 particles, ejected isotropically from Mercury. During the 10\,Myr simulation these particles were removed if either, 1) they collided with one of the 6 massive bodies (including the Sun), or, 2) their apocenter fell below 0.2\,AU (which we considered as safely removed from Mercury's vicinity). Our earlier, analytic treatment assumed coplanarity between the ejecta orbits and Mercury's orbit. This assumption is easily lifted within $N$-body simulations. If the planets and ejecta were assumed coplanar, re-accretion timescales would be deceptively short, owing to the enhanced probability of collision between coplanar orbits as compared to mutually inclined orbits \citep{Farinella1992}. 

We initialize the planetary orbits with similar eccentricities, inclinations and semi-major axes to their modern values, but with uniformly randomized mean anomalies, arguments of pericenter and longitudes of ascending node. The exception is Mercury itself, which we began with mean anomaly, argument of pericenter and longitude of ascending node all set to zero, i.e, the planet began at $y=0$ on our coordinate grid (Figure~\ref{Schematic}). This set-up was simply to aid in prescribing the initial particle orbits relative to Mercury, which we now describe.

With respect to the ejected particles, we adopted a uniform distribution of ejecta angles over 4$\pi$ steradians. Specifically, we chose $N=11$ angles, uniformly spaced in ranges $0< \phi_i\leq2\pi$ and $N-1$ angles from $-1<\cos(\theta_j)<1$, where $\theta_i$ and $\phi_j$ take their usual definitions within a spherical coordinate system centered upon Mercury\footnote{The $\sim6$ degree inclination of Mercury's orbit within our simulation frame will impart a slight misalignment between the $z$-axis of the simulation frame and the $z-$axis of the Mercury-centered frame. However, this offset is not important to the dynamics.} and are defined as
\begin{align}
\phi_i&=\frac{i}{N}2\pi\,\,\,\,\,\,\,\,\,\,\,\,\,&i\in \{1...N\}\nonumber\\
\theta_j&=\arccos\bigg[-1+2\frac{j}{N}\bigg]\,\,\,\,\,\,\,\,\,\,\,\,\,&j\in \{1...N-1\},
\end{align}
yielding $N(N-1)=110$ ejected particles.  

Given the above choices of angles and ejecta velocities, we set up the particles on heliocentric orbits. Having chosen the angles $\theta_i$ and $\phi_j$, we initialize each particle with a position 
\begin{align}
\mathbf{x}_{ij}=x_0\{\sin(\theta_i)\cos(\phi_j),\sin(\theta_i)\sin(\phi_j),\cos(\theta_i)\}
\end{align}
relative to Mercury, such that the initial Heliocentric position is $\mathbf{x}_{ij}+\mathbf{x}_M$ where $\mathbf{x}_M$ is Mercury's initial position. Likewise, the initial velocity of the particle relative to Mercury is given by
\begin{align}
\mathbf{v}_{ij}=v_0\{\sin(\theta_i)\cos(\phi_j),\sin(\theta_i)\sin(\phi_j),\cos(\theta_i)\}.
\end{align}
To ensure that the particle is well clear of Mercury at the beginning of the integration, we set $x_0=0.006\,$AU, which is approximately 4 Hill radiii from Mercury. The value of $v_0$ took either of $\{4,\,10\}\,$kms$^{-1}$. 

\subsection{N-body results}

\begin{figure}[h!]
\centering
\includegraphics[trim=0cm 0cm 0cm 0cm, clip=true,width=1\columnwidth]{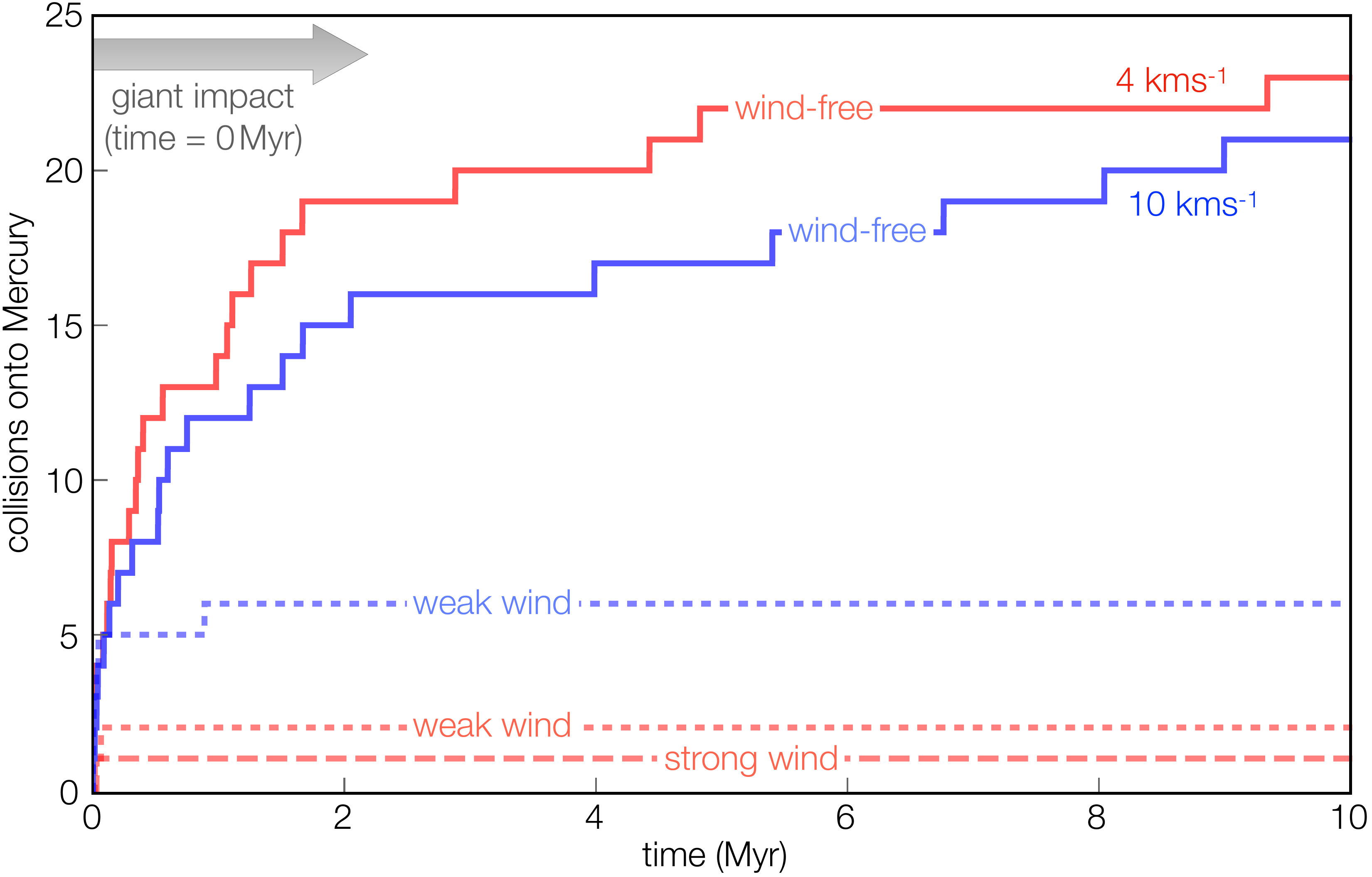}
\caption{The number of particles, from an initial population of 110, that re-impact Mercury subsequent to being launched. Red and blue correspond to launch velocities of 4km/s and 10km/s respectively. Short-dashed lines correspond to drag associated with a wind of $10\dot{M}_\odot$ (``weak wind") and long-dashed lines to winds of $100\dot{M}_\odot$ (``strong wind"). Even moderate winds significantly reduce the number of particles that re-impact Mercury's surface, allowing them to remain unbound from the planet.}
\label{NumSim}
\end{figure}

The cumulative number of collisions onto Mercury within our numerical simulations is presented in Figure~\ref{NumSim}. The wind-free examples agree reasonably well with \citet{Gladman2009}, in that typical re-accretion timescales are 10 million years. Note that we underestimate the number of collisions arising from 4\,kms$^{-1}$ ejections ($\sim20$\,\% vs $\sim 40$\% after 10\,Myr). The reason for this discrepancy likely arises owing to the stochasticity inherent to the problem, together with the smaller number of ejected particles (110) used in the present study, as compared to the 7000 used in \citep{Gladman2009}. For the purposes of the present study, a comparison between the cases with and without wind-induced drag is the primary focus. 

The re-accretion rates occurring under the influence of wind-induced drag were also presented in Figure~\ref{NumSim}. Both $\dot{M}=100\dot{M}_\odot$ (strong wind) and even $\dot{M}=10\dot{M}_\odot$ (weak wind) were sufficient to prevent most re-accretion, with the former allowing only 1 or 2 collisions upon Mercury, as opposed to $\sim20-30$ with no drag. These numerical results confirm the expectation from our analytics that drag timescales corresponding to the early Solar wind prevent most of the re-accretion of material ejected from Mercury's surface. 

In conclusion, within the scope of our analysis, the giant impact origin for Mercury's large core is consistent with the dynamical trajectory taken by material ejected from the primordial Mercury within the early Solar system, permeated by the young Sun's enhanced wind.

\section{Discussion and Conclusions}\label{Discussion}

By way of a combination of semi-analytical theory and numerical simulations, we have shown that the drag associated with the early Sun's wind was likely sufficient to cause orbital migration of cm-sized particles over $<1$\,Myr timescales. Previous work has suggested that a giant impact upon Mercury's progenitor would expel its mantle into heliocentric orbit, where the material condenses into $\sim$\,cm-sized objects. Accordingly, we demonstrated that the addition of the Solar wind, which would have been significantly larger than today at the time of impact, reduces the likelihood that much of this ejected material would re-accrete onto Mercury.

Alternative solutions to the problem of re-accretion have been proposed. The Poynting-Robertson effect, for example, is most analogous to the influence of the Solar wind \citep{Burns1979}. Previous work has shown that the timescale associated with Poynting-Robertson drag is indeed shorter than 10\,Myr for nominal parameters. Criticism of the Poynting-Robertson drag as a solution has included the argument that the optical depth of the purported ring of debris may exceed unity \citep{Gladman2009}. This would increase the timescale of removal, allowing more time for re-accretion. 

The benefit of the Solar wind in this context is that it is potentially 10-100 times stronger, so even if the penetration of Solar wind ions is somewhat reduced by the presence of a dense ring of material, there exist multiple orbital decay times within which to remove the ring. In addition, differential apsidal precession of ring particles would give the ring a non-zero scale height that would reduce the degree of self-shielding. The issue of self-shielding would be worth exploring in follow-up work.

\subsection{Timing of impact}\label{timing}

Age constraints upon the final formation times of the inner terrestrial planets are largely based upon isotopic data of both Earth rocks and Martian meteorites \citep{Raymond2018}. In particular, the radioactive decay system of W-Hf suggests that Earth's formation time spanned tens of millions of years, with its final giant impact perhaps occurring upwards of 100\,Myr after the beginning of planet formation \citep{Kleine2009,Fischer2018}. Mars, in contrast, likely formed substantially earlier, within a few Myr and before the dissipation of the protoplanetary nebula gas \citep{Dauphas2011}. 

Unfortunately, similar direct geochemical constraints are absent for Mercury (and Venus) owing to a lack of Mercurian meteoritic samples. The likely time of Mercury's mantle-stripping impact is therefore poorly constrained. Indirect inferences may be made using the Earth-moon system, which likely formed from a similar giant impact \citep{Canup2001}, hinting that Mercury's final giant impact may have occurred within a similar timeframe; within $\sim100$\,Myr. In principle, a lower limit on the age of Mercury's surface might be deduced from crater size-frequency analysis \citep{Strom2011}. However, an absolute chronology for Mercury's cratering record is largely absent, such that it is difficult to rule out impacts that occured within the first few million years after disk dissipation. 


Uncertainty in the timing of the impact is not critical to our hypothesis in a qualitative sense; ejecta launched within the first $\sim100$\,Myr is uniformly removed from Mercury's orbit. However, the decaying strength of the wind suggests that if the impact occurred earlier than $\sim10$\,Myr, debris would be pushed outward to wider orbits than Mercury, potentially becoming part of Venus or Earth (see Figure~\ref{DragCompare}). This contrasts with debris launched later, which is more likely to simply decay onto the Sun's surface. If the final impact occurred earlier than 10\,Myr, it would suggest that Mercury's formation time is more in-line with Mars \citep{Dauphas2011}. Mars is consistent with a planetary embryo, which had its growth halted, possibly owing to the inward migration of Jupiter \citep{Walsh2011,Batygin2015}. Interestingly, Mercury's suspected initial mass, at 2.25 times its current value \citep{Benz2007} is only about 10\% larger than Mars' current mass. This raises the possibility that Mercury, too, might have been a stranded embryo, later altered during the phase of giant impacts \citep{Raymond2006,Raymond2018}.

Ultimately, there is no specific reason to favor late or early impact scenarios for Mercury, owing to incomplete age-constraints upon its surface. However, large uncertainties likewise exist in the early wind's properties, and so even with accurate ages for Mercury's impact it may still be difficult to say with great confidence whether the debris ended up in the Sun or in the other terrestrial planets. Nevertheless, our modeling here indicates that at all times earlier than 100\,Myr, and subsequent to disk dissipation, the Sun's wind was likely strong enough to remove ejected material from Mercury's orbital vicinity and prevent substantial re-accretion.

\subsection{Size distribution of ejecta}\label{Size}

Throughout this work, we implicitly assumed that the majority of the mass of ejected particles existed as cm-sized spheres. This assumption was chosen in light of thermodynamic arguments presented in \citet{Benz2007}. All timescales computed in our work depend linearly upon particle sizes, and thus it is important to briefly return to the robustness of our assumption of 1\,cm particles. In particular, \citet{Benz2007} followed the thermodynamic evolution of material that is initially shocked to a supercritical fluid state, before subsequently cooling along an isentrope. As it cools, the ejecta eventually becomes sub-critical, with the most energetic material initially condensing to a vapor, while less energetic material first reaches a liquid state \citep{Johnson2012}. A full treatment of the cooling process must apply different approaches for each of these thermodynamic outcomes. The work of \citet{Benz2007} used classical nucleation theory \citep{Raizer1960,Zeldovich2002} to model the highest-energy material, but the lower energy material required a different approach \citep{Grady1982,Melosh1991}.

A more recent treatment of the condensation of ejected material was presented in \citet{Johnson2012}, but within a different impact regime--that of the 10\,km impactor that marked the end of the Cretaceous Period on Earth \citep{Alvarez1980}. Their model consisted of a hemispherical, expanding plume of vapor, out of which solid spherules condensed. The size distribution of condensed material was strongly peaked, with a mean value that scales approximately linearly with the impactor radius. Additionally, the average size of debris exhibited strong variation with respect to the impactor's velocity. Specifically, for an impactor of 1000\,km in diameter, the size of debris ranges from roughly $0.01$\,cm at impact velocities of 15\,km/s, rising to $\sim10$\,cm for 25\,km/s, but falling gradually again to 0.01\,cm at speeds of 50\,km/s.

In the case of Mercury, the impactor's diameter (assuming the single impact case \citealt{Benz2007}) likely exceeded 1000\,km, perhaps approaching $\sim 6000$\,km \citep{Chau2018}. Furthermore, the most likely velocity of the impact has been suggested to lie within the range of 20-30\,kms$^{-1}$ \citep{Benz2007,Chau2018}. Given that a 10\,km impactor produces debris with characteristic size $s=350\mu$m, the larger impactor for Mercury will create correspondingly larger debris. For the expected range of impactor sizes, 1000 to 6000\,km, and for an assumed impact speed of 30 kms$^{-1}$, we can scale up the results from \citet{Johnson2012} to estimate that the mean radius of debris would lie in the range $s\approx3-20$\,cm. These values are roughly consistent with our assumed size (1\,cm), but larger debris sizes are possible, and will depend upon the impact velocity and impactor size.

Despite the uncertainty in particle sizes, we may reframe the problem in terms of the largest particles $s_{max}$ that survive after a time $\tau$ subsequent to the giant impact. Using Equation~(\ref{tdrag}), we find
\begin{align}
s_{max}= \bigg(\frac{\dot{M}}{25\dot{M}_\odot}\bigg)\bigg(\frac{\tau}{\textrm{Myr}}\bigg)\,\textrm{cm},
\end{align}
which suggests that even if the typical size of particles was closer to 10\,cm, as our extrapolations from \citet{Johnson2012} suggest, wind strengths of $100\dot{M}_\odot$ remain sufficient for removal. It is unclear what fraction of the total mass of ejected material would reside in the largest particles, and thus it is worth noting that all ejecta smaller than $s_{\max}$ would also be removed.

Observational constraints upon the remnants of giant impacts may become available in the form of debris disks around other stars \citep{Genda2015}, some of which are already interpreted as indicative of past giant impacts, for example BD +20 307 \citep{Weinberger2010} and HD172555 \citep{Lisse2009,Johnson2012b}. This latter system possesses a population of fine dust ($s\lesssim100$\,$\mu$m) in addition to a population of larger dust inferred to match $Spitzer$ data. There are few direct probes available of cm-sized populations around analogues of the young Solar system, and thus it remains difficult to validate the assumption of cm-sizes using existing data. 

In addition to particle size, it is essential to consider whether an appropriate mass of silicates is removed from the system to account for Mercury's composition. While a full treatment of this problem is beyond the scope of the present investigation, we note that at the high energies required to match present-day Mercury the majority of escaping material is likely to become vaporized \citep{Benz2007,Chau2018}. The debris considered in this study condense from this expanding vapor cloud \citep{Johnson2012}, but eventually the condensation sequence is quenched. Accordingly, an uncertain fraction of the vapor column is converted into solid debris--the rest remains in vapor form and is likely removed by photodissociation on a short timescale (potentially $\lesssim1$\,year \citealt{Johnson2012b}). Accordingly, the influence of Solar wind-induced loss of solids upon Mercury's composition depends critically upon the ratio of vapor relative to condensates.

At least within the regime of a 10\,km impactor, collision velocities of 30\,kms$^{-1}$ appear to lead to a roughly equal split in mass between condensates and vapor \citep{Johnson2012}, but it is unclear whether a 50\% vapor fraction is appropriate for Mercury-like giant collisions. In order to determine this fraction with greater precision, it is necessary in the future to undertake simulations of the giant impact itself, while tracking the debris with an appropriate equation of state for shocked silica \citep{Pierazzo1997,Kraus2012}. In tandem, imposing thermodynamical evolution models such as those of \citet{Johnson2012} on top of SPH simulations \citep{Chau2018}, promises to improve estimates of the particle size distribution. If particle sizes derived from these more involved treatments exceed several metres in radius, then unrealistically large Solar winds may be required to remove the debris, and alternative pathways towards removal of material would be required.

%
%

\subsection{Implications for exoplanets}

Mercury's high density is anomalous in the Solar system, but a growing number of exoplanetary density measurements are facilitating a comparison to exoplanetary analogues. While Mercury-sized planets have been detected around other stars \citep{Barclay2013}, few density estimates exist of this size class \citep{JontofHutter2016}. The only current example, Kepler-138b, is a Mars-sized body with both mass and radius measurements, but does not show enhanced density and furthermore exists around an M-dwarf star \citep{JontofHutter2015}.

If the stellar wind is instrumental in removing debris from giant impacts, we would make the qualitative prediction that closer-in rocky planets are capable of reaching densities that exceed those of more distant planets. This prediction follows from the expectation that subsequent giant impacts selectively remove silicates \citep{Marcus2009,Bonomo2019}, but only if the planet is close enough to the host star for the material to be removed. Such a trend has recently been revealed \citep{Swain2019}, where hotter terrestrial planets reach higher densities than cooler examples. Nevertheless, such a general prediction is confounded by the unknown degree to which stellar winds vary from system to system, and the enhanced impact speeds and reduced re-accretion times expected at shorter orbital periods \citep{Volk2015}.

Extrasolar material in the more exotic form of polluted white dwarfs \citep{Jura2014} provide additional constraints upon the occurrence of Mercury-like planetary compositions. Specifically, white dwarfs present in their spectra evidence of rocky material recently and continuously falling onto their surfaces. It was recently become possible to analyze the composition of this material and determine the Fe:Si ratios and oxygen fugacities they possess \citep{Doyle2019}. Although the parent bodies and dynamical histories of white dwarf pollution remain unidentified, even among this set of cosmochemical measurements Mercury's oxygen fugacity and Fe:Si ratio stand as anomalies.

\subsection{The status of the giant impact hypothesis}

The giant impact scenario for Mercury's formation has existed for decades and taken numerous forms \citep{Chapman1988,Benz2007,Asphaug2014,Chau2018}. In this paper, we have focused upon one specific concern with this model, and that is the likelihood that any material blasted off Mercury is likely to re-accrete over a short timescale. We have shown that the Solar wind constitutes a viable pathway to removing this ejecta, and thereby alleviating this problem with the giant impact hypothesis. 

The \textit{MESSENGER} mission has afforded a higher level view of the evidence than merely the Si:Fe mass ratio \citep{Solomon2018}. In addition to Mercury's large iron content, the planet's surface was revealed to possess an abnormally low oxygen fugacity with respect to the other terrestrial planets \citep{Nittler2011,McCubbin2012}. These chemical constraints are difficult to compare with the giant impact framework, in part owing to computational limitations related to following individual chemical species within the Smoothed-Particle Hydrodynamics simulations typically applied to giant impact investigations \citep{Lock2017,Chau2018}, and furthermore due to the difficulty in addressing the degree of equilibration between silicates and iron within each body \citep{Dahl2010}. 

An additional revelation, that the surface of Mercury is volatile-rich, was once considered as inconsistent with a giant impact model \citep{Peplowski2011}. The idea that giant impacts preferentially remove volatiles has typically been motivated by low volatile contents of the Moon \citep{Stewart2016}, itself an outcome of a giant impact. However, the volatile depletion of the Moon remains difficult to replicate within giant impact simulations, requiring fractionation between a moon-forming disk and the proto-Earth \citep{Canup2015}. In other words, the volatile depletion seen on the Moon occurred as a result of it being the remnant of the impact, forming out of circumplanetary material that experienced volatile fractionation \citep{Lock2016,Lock2017,Lock2018,Nakajima2018}. With respect to Mercury's high volatile content, then, it is more appropriate to compare Mercury to the Earth and not to the Moon \citep{Stewart2016}.

A final point regarding volatiles is that giant impacts constitute a robust expectation during the latter stages of terrestrial planet formation generally, across a range of models \citep{Stewart2016,Raymond2018}. Therefore, it is likely that Venus, Earth and even Mars experienced giant collisions and yet exhibit similar volatile contents to Mercury \citep{Ebel2018}. Accordingly, the high volatile content of Mercury no longer appears inconsistent with the planet having experiencing a giant impact at the end of its formation epoch.

The feasibility of the giant impact hypothesis for the origin of Mercury's anomalously large core remains contentious. Nevertheless, the problem of re-accretion of ejected material becomes significantly reduced when the influence of the Solar wind is taken into account. Thus, in early planetary systems, the stellar wind may play an instrumental, yet under-appreciated role in the final architectures of planetary systems. Within this context, Mercury's anomalous density stands as the nearest, and most familiar example of the role of the Sun's wind in the early Solar system. 

\begin{acknowledgments}
 C.S thanks Gregory Laughlin, Ravit Helled, Brett Gladman and Sarah Stewart for useful discussions, along with the 51 Pegasi b Heising-Simons Foundation grant for their generous support. F. C. A acknowledges support from the  NASA Exoplanets Research Program. We acknowledge the thoughtful comments of two anonymous reviewers, whose input substantially improved the quality of the manuscript.
\end{acknowledgments}


\begin{thebibliography}{}

\bibitem[Alexander et al.(2001)]{Alexander2001}Alexander, C. O. D., Boss, A. P., \& Carlson, R. W. (2001). The early evolution of the inner solar system: A meteoritic perspective. Science, 293(5527), 64-68.

\bibitem[Alvarez et al.(1980)]{Alvarez1980}Alvarez, L. W., Alvarez, W., Asaro, F., \& Michel, H. V. (1980). Extraterrestrial cause for the Cretaceous--Tertiary extinction. Science, 208(4448), 1095-1108.

\bibitem[Armitage(2011)]{Armitage2011}Armitage, P. J. (2011). Dynamics of protoplanetary disks. Annual Review of Astronomy and Astrophysics, 49.

\bibitem[Ash et al.(1971)]{Ash1971}Ash, M. E., Shapiro, I. I., \& Smith, W. B. (1971). The system of planetary masses. Science, 174(4009), 551-556.

\bibitem[Asphaug et al.(2006)]{Asphaug2006}Asphaug, E., Agnor, C. B., \& Williams, Q. (2006). Hit-and-run planetary collisions. Nature, 439(7073), 155.

\bibitem[Asphaug \& Reufer(2014)]{Asphaug2014}Asphaug, E., \& Reufer, A. (2014). Mercury and other iron-rich planetary bodies as relics of inefficient accretion. Nature Geoscience, 7(8), 564.

\bibitem[Barclay et al.(2013)]{Barclay2013}Barclay, T., Rowe, J. F., Lissauer, J. J., Huber, D., Fressin, F., Howell, S. B., ... \& Marcy, G. W. (2013). A sub-Mercury-sized exoplanet. Nature, 494(7438), 452.

\bibitem[Batygin \& Laughlin(2015)]{Batygin2015}Batygin, K., \& Laughlin, G. (2015). Jupiter's decisive role in the inner Solar System?s early evolution. Proceedings of the National Academy of Sciences, 112(14), 4214-4217.

\bibitem[Bellan(2008)]{Bellan2008}Bellan, P. M. (2008). Fundamentals of plasma physics. Cambridge University Press.

\bibitem[Benz et al.(1988)]{Benz1988}Benz, W., Slattery, W. L., \& Cameron, A. G. W. (1988). Collisional stripping of Mercury's mantle. Icarus, 74(3), 516-528.


\bibitem[Benz et al.(2007)]{Benz2007}Benz, W., Anic, A., Horner, J., \& Whitby, J. A. (2007). The Origin of Mercury. Space Sci Rev, 132, 189-202.


\bibitem[Bonomo et al.(2019)]{Bonomo2019}Bonomo, A. S., Zeng, L., Damasso, M., Leinhardt, Z. M., Justesen, A. B., Lopez, E., ... \& Corsaro, E. (2019). A giant impact as the likely origin of different twins in the Kepler-107 exoplanet system. Nature Astronomy, 3(5), 416.

\bibitem[Borucki(2016)]{Borucki2016}Borucki, W. J. (2016). KEPLER Mission: development and overview. Reports on Progress in Physics, 79(3), 036901.

\bibitem[Bouvier et al.(2014)]{Bouvier2014}Bouvier, J., Matt, S. P., Mohanty, S., Scholz, A., Stassun, K. G., \& Zanni, C. (2014). Angular momentum evolution of young low-mass stars and brown dwarfs: observations and theory. Protostars and Planets VI, 433, 94.


\bibitem[Burns(1976)]{Burns1976}Burns, J. A. (1976). Elementary derivation of the perturbation equations of celestial mechanics. American Journal of Physics, 44(10), 944-949.

\bibitem[Burns et al.(1979)]{Burns1979}Burns, J. A., Lamy, P. L., \& Soter, S. (1979). Radiation forces on small particles in the solar system. Icarus, 40(1), 1-48.

\bibitem[Cameron(1985)]{Cameron1985}Cameron, A. G. W. (1985). The partial volatilization of Mercury. Icarus, 64(2), 285-294.

\bibitem[Canup \& Asphaug(2001)]{Canup2001}Canup, R. M., \& Asphaug, E. (2001). Origin of the Moon in a giant impact near the end of the Earth's formation. Nature, 412(6848), 708.

\bibitem[Canup et al.(2015)]{Canup2015}Canup, R. M., Visscher, C., Salmon, J., \& Fegley Jr, B. (2015). Lunar volatile depletion due to incomplete accretion within an impact-generated disk. Nature geoscience, 8(12), 918-921.

\bibitem[Canup(2005)]{Canup2005}Canup, R. M. (2005). A giant impact origin of Pluto-Charon. Science, 307(5709), 546-550.

\bibitem[Carolan et al.(2019)]{Carolan2019}Carolan, S., Vidotto, A. A., Loesch, C., \& Coogan, P. (2019). The evolution of Earth's magnetosphere during the solar main sequence. Monthly Notices of the Royal Astronomical Society, 489(4), 5784-5801..


\bibitem[Chapman(1988)]{Chapman1988}Chapman, C. R. (1988). Mercury: Introduction to an end-member planet. Mercury, 1-23.

\bibitem[Chau et al.(2018)]{Chau2018}Chau, A., Reinhardt, C., Helled, R., \& Stadel, J. (2018). Forming Mercury by giant impacts. The Astrophysical Journal, 865(1), 35.


\bibitem[Citron et al.(2015)]{Citron2015}Citron, R. I., Genda, H., \& Ida, S. (2015). Formation of Phobos and Deimos via a giant impact. Icarus, 252, 334-338.


\bibitem[Cohen \& Drake(2014)]{Cohen2014}Cohen, O., \& Drake, J. J. (2014). A grid of MHD models for stellar mass loss and spin-down rates of solar analogs. The Astrophysical Journal, 783(1), 55.


%

\bibitem[Cranmer \& Saar(2011)]{Cranmer2011}Cranmer, S. R., \& Saar, S. H. (2011). Testing a predictive theoretical model for the mass loss rates of cool stars. The Astrophysical Journal, 741(1), 54.

\bibitem[\'Cuk \& Stewart(2012)]{Cuk2012}\'Cuk, M., \& Stewart, S. T. (2012). Making the Moon from a fast-spinning Earth: a giant impact followed by resonant despinning. science, 338(6110), 1047-1052.

\bibitem[Dahl \& Stevenson(2010)]{Dahl2010}Dahl, T. W., \& Stevenson, D. J. (2010). Turbulent mixing of metal and silicate during planet accretion--And interpretation of the Hf--W chronometer. Earth and Planetary Science Letters, 295(1-2), 177-186.

\bibitem[Dauphas \& Pourmand(2011)]{Dauphas2011}Dauphas, N., \& Pourmand, A. (2011). Hf--W--Th evidence for rapid growth of Mars and its status as a planetary embryo. Nature, 473(7348), 489-492.

\bibitem[Donati \& Landstreet(2009)]{Donati2009}Donati, J. F., \& Landstreet, J. D. (2009). Magnetic fields of nondegenerate stars. Annual Review of Astronomy and Astrophysics, 47, 333-370.


\bibitem[Doyle et al.(2019)]{Doyle2019}Doyle, A. E., Young, E. D., Zuckerman, B., Klein, B., \& Schlichting, H. E. (2019). Doyle, A. E., Young, E. D., Klein, B., Zuckerman, B., \& Schlichting, H. E. (2019). Oxygen fugacities of extrasolar rocks: Evidence for an Earth-like geochemistry of exoplanets. Science, 366(6463), 356-359.


\bibitem[Ebel \& Stewart(2018)]{Ebel2018}Ebel, D. S., \& Stewart, S. T. (2018). The elusive origin of Mercury. In Mercury: The view after \textit{MESSENGER} (Vol. 21). Cambridge University Press.



\bibitem[Farinella et al.(1992)]{Farinella1992}Farinella, Paolo, and Donald R. Davis. Collision rates and impact velocities in the main asteroid belt. Icarus 97.1 (1992): 111-123.

\bibitem[Fernandez et al.(2006)]{Fernandez2006}Fern\'andez, R., Brandeker, A., \& Wu, Y. (2006). Braking the gas in the $\beta$ Pictoris disk. The Astrophysical Journal, 643(1), 509.


\bibitem[Fischer \& Nimmo(2018)]{Fischer2018}Fischer, R. A., \& Nimmo, F. (2018). Effects of core formation on the Hf-W isotopic composition of the Earth and dating of the Moon-forming impact. Earth and Planetary Science Letters, 499, 257-265.

\bibitem[Genda et al.(2015)]{Genda2015}Genda, H., Kobayashi, H., \& Kokubo, E. (2015). Warm debris disks produced by giant impacts during terrestrial planet formation. The Astrophysical Journal, 810(2), 136.

\bibitem[Gladman et al.(1996)]{Gladman1996}Gladman, B. J., Burns, J. A., Duncan, M., Lee, P., \& Levison, H. F. (1996). The exchange of impact ejecta between terrestrial planets. Science, 271(5254), 1387-1392.

\bibitem[Gladman \& Coffey(2009)]{Gladman2009}Gladman, B., \& Coffey, J. (2009). Mercurian impact ejecta: Meteorites and mantle. Meteoritics \& Planetary Science, 44(2), 285-291.

\bibitem[Gomes et al.(2005)]{Gomes2005}Gomes, R., Levison, H. F., Tsiganis, K., \& Morbidelli, A. (2005). Origin of the cataclysmic Late Heavy Bombardment period of the terrestrial planets. Nature, 435(7041), 466.

\bibitem[Grady(1982)]{Grady1982}Grady, D. E. (1982). Local inertial effects in dynamic fragmentation. Journal of Applied Physics, 53(1), 322-325.

\bibitem[Hartmann \& Davis(1975)]{Hartmann1975}Hartmann, W. K., \& Davis, D. R. (1975). Satellite-sized planetesimals and lunar origin. Icarus, 24(4), 504-515.

\bibitem[Hartmann \& MacGregor(1982)]{Hartmann1982}Hartmann, L. M., \& MacGregor, K. B. (1982). Protostellar mass and angular momentum loss. The Astrophysical Journal, 259, 180-192.

\bibitem[Hartmann et al.(2016)]{Hartmann2016}Hartmann, L., Herczeg, G., \& Calvet, N. (2016). Accretion onto pre-main-sequence stars. Annual Review of Astronomy and Astrophysics, 54, 135-180.

\bibitem[Hedman(2018)]{Hedman2018}Hedman, M. M. (2018). An Introduction to Planetary Ring Dynamics. Planetary Ring Systems. Properties, Structure, and Evolution, 30-48.

\bibitem[Hubbard(2014)]{Hubbard2014}Hubbard, A. (2014). Explaining Mercury's density through magnetic erosion. Icarus, 241, 329-335.

\bibitem[Hyodo et al.(2017)]{Hyodo2017}Hyodo, R., Genda, H., Charnoz, S., \& Rosenblatt, P. (2017). On the impact origin of Phobos and Deimos. I. Thermodynamic and physical aspects. The Astrophysical Journal, 845(2), 125.

\bibitem[Jardine \& Collier Cameron(2018)]{Jardine2018}Jardine, M., \& Collier Cameron, A. (2018). Slingshot prominences: nature's wind gauges. Monthly Notices of the Royal Astronomical Society, 482(3), 2853-2860.

\bibitem[Johnson \& Melosh(2012)]{Johnson2012}Johnson, B. C., \& Melosh, H. J. (2012). Formation of spherules in impact produced vapor plumes. Icarus, 217(1), 416-430.

\bibitem[Johnson et al.(2012)]{Johnson2012b}Johnson, B. C., Lisse, C. M., Chen, C. H., Melosh, H. J., Wyatt, M. C., Thebault, P., ... \& Morlok, A. (2012). A self-consistent model of the circumstellar debris created by a giant hypervelocity impact in the HD 172555 system. The Astrophysical Journal, 761(1), 45.

\bibitem[Jontof-Hutter et al.(2015)]{JontofHutter2015}Jontof-Hutter, D., Rowe, J. F., Lissauer, J. J., Fabrycky, D. C., \& Ford, E. B. (2015). The mass of the Mars-sized exoplanet Kepler-138 b from transit timing. Nature, 522(7556), 321.

\bibitem[Jontof-Hutter et al.(2016)]{JontofHutter2016}Jontof-Hutter, D., Ford, E. B., Rowe, J. F., Lissauer, J. J., Fabrycky, D. C., Van Laerhoven, C., ... \& Mazeh, T. (2016). Secure mass measurements from transit timing: 10 Kepler exoplanets between 3 and 8 M$_\oplus$ with diverse densities and incident fluxes. The Astrophysical Journal, 820(1), 39.


\bibitem[Jura \& Young(2014)]{Jura2014}Jura, M., \& Young, E. D. (2014). Extrasolar cosmochemistry. Annual Review of Earth and Planetary Sciences, 42, 45-67.

\bibitem[Kasper et al.(2019)]{Kasper2019}Kasper, J. C., Bale, S. D., Belcher, J. W., Berthomier, M., Case, A. W., Chandran, B. D. G., ... \& Halekas, J. S. (2019). Alfv\'enic velocity spikes and rotational flows in the near-Sun solar wind. Nature, 576(7786), 228-231.

\bibitem[Kleine et al.(2009)]{Kleine2009}Kleine, T., Touboul, M., Bourdon, B., Nimmo, F., Mezger, K., Palme, H., ... \& Halliday, A. N. (2009). Hf-W chronology of the accretion and early evolution of asteroids and terrestrial planets. Geochimica et Cosmochimica Acta, 73(17), 5150-5188.

\bibitem[Kraus et al.(2012)]{Kraus2012}Kraus, R. G., Stewart, S. T., Swift, D. C., Bolme, C. A., Smith, R. F., Hamel, S., ... \& Collins, G. W. (2012). Shock vaporization of silica and the thermodynamics of planetary impact events. Journal of Geophysical Research: Planets, 117(E9).

\bibitem[Lamers \& Cassinelli(1999)]{Lamers1999}Lamers, H. J., Cassinelli, J. P., \& Cassinelli, J. (1999). Introduction to stellar winds. Cambridge university press.



\bibitem[Lewis(1973)]{Lewis1973}Lewis, J. S. (1973). Chemistry of the planets. Annual Review of Physical Chemistry, 24(1), 339-351.

\bibitem[Lisse et al.(2009)]{Lisse2009}Lisse, C. M., Chen, C. H., Wyatt, M. C., Morlok, A., Song, I., Bryden, G., \& Sheehan, P. (2009). Abundant circumstellar silica dust and SiO gas created by a giant hypervelocity collision in the? 12 Myr HD172555 system. The Astrophysical Journal, 701(2), 2019.

\bibitem[Lock et al.(2016)]{Lock2016}Lock, S. J., Stewart, S. T., Petaev, M. I., Leinhardt, Z. M., Mace, M., Jacobsen, S. B., \& \'Cuk, M. (2016). A new model for lunar origin: equilibration with Earth beyond the hot spin stability limit. In Lunar and Planetary Science Conference (Vol. 47, p. 2881).

\bibitem[Lock \& Stewart(2017)]{Lock2017}Lock, S. J., \& Stewart, S. T. (2017). The structure of terrestrial bodies: Impact heating, corotation limits, and synestias. Journal of Geophysical Research: Planets, 122(5), 950-982.

\bibitem[Lock et al.(2018)]{Lock2018}Lock, S. J., Stewart, S. T., Petaev, M. I., Leinhardt, Z., Mace, M. T., Jacobsen, S. B., \& Cuk, M. (2018). The origin of the Moon within a terrestrial synestia. Journal of Geophysical Research: Planets, 123(4), 910-951.

\bibitem[Lodders et al.(2009)]{Lodders2009}Lodders, K., Palme, H., \& Gail, H. P. (2009). 4.4 Abundances of the elements in the Solar System. In Solar system (pp. 712-770). Springer, Berlin, Heidelberg.

\bibitem[Lovelace et al.(2008)]{Lovelace2008}Lovelace, R. V. E., Romanova, M. M., \& Barnard, A. W. (2008). Planet migration and disc destruction due to magneto-centrifugal stellar winds. Monthly Notices of the Royal Astronomical Society, 389(3), 1233-1239.


\bibitem[Marcus et al.(2009)]{Marcus2009}Marcus, R. A., Stewart, S. T., Sasselov, D., \& Hernquist, L. (2009). Collisional stripping and disruption of super-Earths. The Astrophysical Journal Letters, 700(2), L118.


\bibitem[Maeder(2008)]{Maeder2008}Maeder, A. (2008). Physics, formation and evolution of rotating stars. Springer Science \& Business Media.

\bibitem[Matt \& Pudritz(2008)]{Matt2008}Matt, S., \& Pudritz, R. E. (2008). Accretion-powered stellar winds. II. Numerical solutions for stellar wind torques. The Astrophysical Journal, 678(2), 1109.

\bibitem[McCubbin et al.(2012)]{McCubbin2012}McCubbin, F. M., Riner, M. A., Vander Kaaden, K. E., \& Burkemper, L. K. (2012). Is Mercury a volatile-rich planet? Geophysical Research Letters, 39(9).

\bibitem[Melosh \& Vickery(1991)]{Melosh1991}Melosh, H. J., \& Vickery, A. M. (1991). Melt droplet formation in energetic impact events. Nature, 350(6318), 494.

\bibitem[Morbidelli et al.(2012)]{Morbidelli2012}Morbidelli, A., Tsiganis, K., Batygin, K., Crida, A., \& Gomes, R. (2012). Explaining why the uranian satellites have equatorial prograde orbits despite the large planetary obliquity. Icarus, 219(2), 737-740.


\bibitem[Mukai \& Yamamoto(1982)]{Mukai1982}Mukai, T., \& Yamamoto, T. (1982). Solar wind pressure on interplanetary dust. Astronomy and Astrophysics, 107, 97-100.

\bibitem[Murray \& Dermott(1999)]{Murray1999}Murray, C. D., \& Dermott, S. F. (1999). Solar system dynamics. Cambridge university press.

\bibitem[Nakajima \& Stevenson(2018)]{Nakajima2018}Nakajima, M., \& Stevenson, D. J. (2018). Inefficient volatile loss from the Moon-forming disk: Reconciling the giant impact hypothesis and a wet Moon. Earth and Planetary Science Letters, 487, 117-126.


\bibitem[Nittler et al.(2011)]{Nittler2011}Nittler, L. R., Starr, R. D., Weider, S. Z., McCoy, T. J., Boynton, W. V., Ebel, D. S., ... \& Lawrence, D. J. (2011). The major-element composition of Mercury's surface from \textit{MESSENGER} X-ray spectrometry. Science, 333(6051), 1847-1850.

\bibitem[\'O Fionnag\'ain \& Vidotto(2018)]{Fion2018}\'O Fionnag\'ain, D., \& Vidotto, A. A. (2018). The solar wind in time: a change in the behaviour of older winds? Monthly Notices of the Royal Astronomical Society, 476(2), 2465-2475.

\bibitem[Parker(1965)]{Parker1965}Parker, E. N. (1965). Dynamical theory of the solar wind. Space Science Reviews, 4(5-6), 666-708.

\bibitem[Peplowski et al.(2011)]{Peplowski2011}Peplowski, P. N., Evans, L. G., Hauck, S. A., McCoy, T. J., Boynton, W. V., Gillis-Davis, J. J., ... \& McNutt, R. L. (2011). Radioactive elements on Mercury's surface from \textit{MESSENGER}: Implications for the planet's formation and evolution. Science, 333(6051), 1850-1852.

\bibitem[Phillips et al.(1995)]{Phillips1995}Phillips, J. L., Bame, S. J., Barnes, A., Barraclough, B. L., Feldman, W. C., Goldstein, B. E., ... \& Suess, S. T. (1995). Ulysses solar wind plasma observations from pole to pole. Geophysical research letters, 22(23), 3301-3304.

\bibitem[Pierazzo et al.(1997)]{Pierazzo1997}Pierazzo, E., Vickery, A. M., \& Melosh, H. J. (1997). A reevaluation of impact melt production. Icarus, 127(2), 408-423.

\bibitem[Plavchan et al.(2009)]{Plavchan2009}Plavchan, P., Werner, M. W., Chen, C. H., Stapelfeldt, K. R., Su, K. Y. L., Stauffer, J. R., \& Song, I. (2009). New debris disks around young, low-mass stars discovered with the Spitzer Space Telescope. The Astrophysical Journal, 698(2), 1068.


\bibitem[Raizer(1960)]{Raizer1960}Raizer, Y. P. (1960). Condensation of a cloud of vaporized matter expanding in vacuum. SOVIET PHYSICS JETP-USSR, 10(6), 1229-1235.

\bibitem[Raymond et al.(2004)]{Raymond2004}Raymond, S. N., Quinn, T., \& Lunine, J. I. (2004). Making other earths: dynamical simulations of terrestrial planet formation and water delivery. Icarus, 168(1), 1-17.

\bibitem[Raymond et al.(2006)]{Raymond2006}Raymond, S. N., Quinn, T., \& Lunine, J. I. (2006). High-resolution simulations of the final assembly of Earth-like planets I. Terrestrial accretion and dynamics. Icarus, 183(2), 265-282.

\bibitem[Raymond et al.(2018)]{Raymond2018}Raymond, S. N., Izidoro, A., \& Morbidelli, A. (2018). Solar System Formation in the Context of Extra-Solar Planets. arXiv preprint arXiv:1812.01033.

\bibitem[Rein \& Liu(2012)]{Rein2012}Rein, H., \& Liu, S. F. (2012). REBOUND: an open-source multi-purpose N-body code for collisional dynamics. Astronomy \& Astrophysics, 537, A128.

\bibitem[Rein \& Spiegel(2014)]{Rein2014}Rein, H., \& Spiegel, D. S. (2014). IAS15: a fast, adaptive, high-order integrator for gravitational dynamics, accurate to machine precision over a billion orbits. Monthly Notices of the Royal Astronomical Society, 446(2), 1424-1437.

\bibitem[Rein \& Tamayo(2015)]{Rein2015}Rein, H., \& Tamayo, D. (2015). WHFAST: a fast and unbiased implementation of a symplectic Wisdom-Holman integrator for long-term gravitational simulations. Monthly Notices of the Royal Astronomical Society, 452(1), 376-388.


\bibitem[Righter et al.(2006)]{Righter2006}Righter, K., Drake, M. J., \& Scott, E. R. D. (2006). Compositional relationships between meteorites and terrestrial planets. Meteorites and the early solar system II, 943, 803-828.

\bibitem[Safronov(1966)]{Safronov1966}Safronov, V. S. (1966). Sizes of the largest bodies falling onto the planets during their formation. Soviet Astronomy, 9, 987-991.

\bibitem[Shu et al.(1987)]{Shu1987}Shu, F. H., Adams, F. C., \& Lizano, S. (1987). Star formation in molecular clouds: observation and theory. Annual review of astronomy and astrophysics, 25(1), 23-81

\bibitem[Siess et al.(2000)]{Siess2000}Siess, L., Dufour, E., \& Forestini, M. (2000). An internet server for pre-main sequence tracks of low-and intermediate-mass stars. Astron. Astrophys, 358, 593-599.

\bibitem[Solomon(2003)]{Solomon2003}Solomon, S. C. (2003). Mercury: the enigmatic innermost planet. Earth and Planetary Science Letters, 216(4), 441-455.

\bibitem[Solomon et al.(2018)]{Solomon2018}Solomon, S. C., Nittler, L. R., \& Anderson, B. J. (Eds.). (2018). Mercury: The view after \textit{MESSENGER} (Vol. 21). Cambridge University Press.

\bibitem[Spalding(2018)]{Spalding2018}Spalding, C. (2018). The Primordial Solar Wind as a Sculptor of Terrestrial Planet Formation. The Astrophysical Journal Letters, 869(1), L17.

\bibitem[Stewart et al.(2016)]{Stewart2016}Stewart, S. T., Lock, S. J., Petaev, M. I., Jacobsen, S. B., Sarid, G., Leinhardt, Z. M., ... \& Humayun, M. (2016, March). Mercury impact origin hypothesis survives the volatile crisis: Implications for terrestrial planet formation. In Lunar and Planetary Science Conference (Vol. 47, p. 2954).

\bibitem[Strom et al.(2011)]{Strom2011}Strom, R. G., Banks, M. E., Chapman, C. R., Fassett, C. I., Forde, J. A., Head III, J. W., ... \& Solomon, S. C. (2011). Mercury crater statistics from \textit{MESSENGER} flybys: Implications for stratigraphy and resurfacing history. Planetary and Space Science, 59(15), 1960-1967.

\bibitem[Svetsov(2011)]{Svetsov2011}Svetsov, V. (2011). Cratering erosion of planetary embryos. Icarus, 214(1), 316-326.

\bibitem[Swain et al.(2019)]{Swain2019}Swain, M. R., Estrela, R., Sotin, C., Roudier, G. M., \& Zellem, R. T. (2019). Two Terrestrial Planet Families with Different Origins. The Astrophysical Journal, 881(2), 117.

\bibitem[Tamayo et al.(2019)]{Tamayo2019}Tamayo, D., Rein, H., Shi, P., \& Hernandez, D. M. (2019). REBOUNDx: A Library for Adding Conservative and Dissipative Forces To Otherwise Symplectic N-body Integrations. Monthly Notices of the Royal Astronomical Society.

\bibitem[Urey(1951)]{Urey1951}Urey, H. C. (1951). The origin and development of the earth and other terrestrial planets. Geochimica et Cosmochimica Acta, 1(4-6), 209-277.

\bibitem[Van der Holst et al.(2014)]{vanderHolst2014}van der Holst, B., Sokolov, I. V., Meng, X., Jin, M., Manchester IV, W. B., T\'oth, G., \& Gombosi, T. I. (2014). Alfvén wave solar model (AWSoM): coronal heating. The Astrophysical Journal, 782(2), 81.

\bibitem[Vidotto et al.(2014)]{Vidotto2014}Vidotto, A. A., Gregory, S. G., Jardine, M., Donati, J. F., Petit, P., Morin, J., ... \& Marsden, S. (2014). Stellar magnetism: empirical trends with age and rotation. Monthly Notices of the Royal Astronomical Society, 441(3), 2361-2374.

\bibitem[Volk \& Gladman(2015)]{Volk2015}Volk, K., \& Gladman, B. (2015). Consolidating and crushing exoplanets: did it happen here?. The Astrophysical Journal Letters, 806(2), L26.

\bibitem[Walsh et al.(2011)]{Walsh2011}Walsh, K. J., Morbidelli, A., Raymond, S. N., O'Brien, D. P., \& Mandell, A. M. (2011). A low mass for Mars from Jupiter?s early gas-driven migration. Nature, 475(7355), 206-209.

\bibitem[Weber \& Davis(1967)]{Weber1967}Weber, E. J., \& Davis Jr, L. (1967). The angular momentum of the solar wind. The Astrophysical Journal, 148, 217-227.

\bibitem[Weidenschilling(1978)]{Weidenschilling1978}Weidenschilling, S. J. (1978). Iron/silicate fractionation and the origin of Mercury. Icarus, 35(1), 99-111.

\bibitem[Weinberger et al.(2010)]{Weinberger2010}Weinberger, A. J., Becklin, E. E., Song, I., \& Zuckerman, B. (2010). The absence of cold dust and the mineralogy and origin of the warm dust encircling BD+ 20 307. The Astrophysical Journal, 726(2), 72.

\bibitem[Wetherill(1985)]{Wetherill1985}Wetherill, G. W. (1985). Occurrence of giant impacts during the growth of the terrestrial planets. Science, 228(4701), 877-879.

\bibitem[Wisdom \& Holman(1991)]{Wisdom1991}Wisdom, J., \& Holman, M. (1991). Symplectic maps for the n-body problem. The Astronomical Journal, 102, 1528-1538.

\bibitem[White \& Hillenbrand(2004)]{White2004}White, R. J., \& Hillenbrand, L. A. (2004). On the evolutionary status of class I stars and Herbig-Haro energy sources in Taurus-Auriga. The Astrophysical Journal, 616(2), 998.

\bibitem[Wood et al.(2014)]{Wood2014}Wood, B. E., Müller, H. R., Redfield, S., \& Edelman, E. (2014). Evidence for a weak wind from the young Sun. The Astrophysical Journal Letters, 781(2), L33.

\bibitem[Wurm et al.(2013)]{Wurm2013}Wurm, G., Trieloff, M., \& Rauer, H. (2013). Photophoretic separation of metals and silicates: The formation of Mercury-like planets and metal depletion in chondrites. The Astrophysical Journal, 769(1), 78.

\bibitem[Zel'dovich \& Raizer(2002)]{Zeldovich2002}Zel'Dovich, Y. B., \& Raizer, Y. P. (2002). Physics of shock waves and high-temperature hydrodynamic phenomena. Courier Corporation.



\end{thebibliography}
\end{document}